\documentclass[prd,twoside,a4paper]{revtex4}

\usepackage{amsmath,amssymb}
\usepackage{graphicx,tabularx}
\usepackage{times}

\newcommand{\D}{\displaystyle}
\newcommand{\qb}{\bar{q}}
\newcommand{\fb}{\bar{f}}
\newcommand{\MX}{\overline{|\mathcal{M}|^2}}
\newcommand{\sla}[1]{/\!\!\!\!#1}
\newcommand\one{\leavevmode\hbox{\small1\normalsize\kern-0.33em1}}

\def\eg{{\sl e.g.} \,}
\def\ie{{\sl i.e.} \,}

\begin{document}

\title{LHC Phenomenology for Physics Hunters}

\author{Tilman Plehn}
\affiliation{SUPA, School of Physics and Astronomy, 
             University of Edinburgh, Scotland}

\begin{abstract}
Welcome to the 2008 TASI lectures on the exciting topic of `tools and
technicalities' (original title). Technically, LHC physics is really
all about perturbative QCD in signals or backgrounds. Whenever we look
for interesting signatures at the LHC we get killed by QCD. Therefore,
I will focus on QCD issues which arise for example in Higgs searches
or exotics searches at the LHC, and ways to tackle them nowadays. In
the last section you will find a few phenomenological discussions, for
example on missing energy or helicity amplitudes.

\end{abstract}

\maketitle
\bigskip\bigskip\bigskip\bigskip

\tableofcontents 
\newpage

\section{LHC Phenomenology}

When we think about signal or background processes at the LHC the
first quantity we compute is the total number of events we would
expect at the LHC in a given time interval. This number of events is
the product of the hadronic (\ie proton--proton) LHC luminosity
measured in inverse femtobarns and the total production cross section
measured in femtobarns. A typical year of LHC running could deliver
around 10 inverse femtobarns per year in the first few years and three
to ten times that later. People who build the actual collider do not
use these kinds of units, but for phenomenologists they work better
than something involving seconds and square meters, because what we
typically need is a few interesting events corresponding to a few
femtobarns of data. So here are a few key numbers and their orders of
magnitude for typical signals:
\begin{equation}
N_{\rm events} = \sigma_{\rm tot} \cdot \mathcal{L} \qquad \quad
\mathcal{L} = 10 \cdots 300 \, \text{fb}^{-1} \qquad \quad
\sigma_{\rm tot} = 1 \cdots 10^4 \, \text{fb} 
\end{equation}
\bigskip

Just in case my colleagues have not told you about it: there are two
kinds of processes at the LHC. The first involves all particles which
we know and love, like old-fashioned electrons or slightly more
modern $W$ and $Z$ bosons or most recently top quarks. These processes
we call
\underline{backgrounds} and find annoying. They are described by QCD, 
which means QCD is the theory of the evil. Top quarks have an
interesting history, because when I was a graduate student they still
belonged to the second class of processes, the
\underline{signals}. These typically involve particles we have not
seen before. Such states are unfortunately mostly produced in QCD
processes as well, so QCD is not entirely evil. If we see such
signals, someone gets a call from Stockholm, shakes hands with the
king of Sweden, and the corresponding processes instantly turn into
backgrounds.

\begin{figure}[t]
\begin{center}
\includegraphics[width=8cm,angle=0]{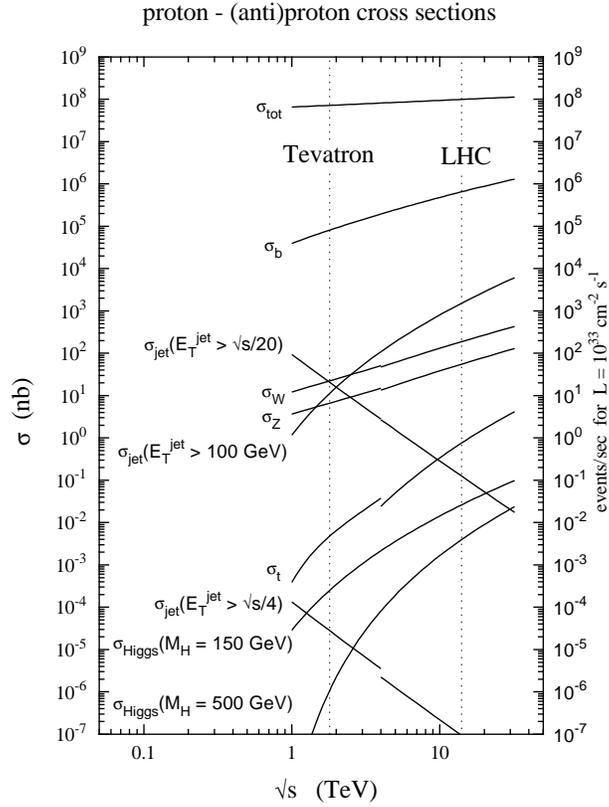}
\end{center}
\vspace*{-0.5cm}
\caption{Production rates for different signal and background
  processes at hadron colliders. The discontinuity is due to the
  Tevatron being a proton--antiproton collider while the LHC is a
  proton--proton collider. The two colliders correspond to the
  $x$--axis values of 2~TeV and 14~TeV. Figure borrowed from CMS.}
\label{fig:lhc}
\end{figure}

The main problem at any collider is that signals are much more rare
that background, so we have to dig our signal events out of a much
larger number of background events. This is what most of this lecture
will be about. Just to give you a rough idea, have a look at
Fig.~\ref{fig:lhc}: at the LHC the production cross section for two
bottom quarks at the LHC is larger than $10^5$~nb or $10^{11}$~fb and
the typical production cross section for $W$ or $Z$ boson ranges
around 200~nb or $2 \times 10^8$~fb. Looking at signals, the
production cross sections for a pair of 500~GeV gluinos is $4 \times
10^4$~fb and the Higgs production cross section can be as big as $2
\times 10^5$~fb. When we want to extract such signals out of
comparably huge backgrounds we need to describe these backgrounds with
an incredible precision. Strictly speaking, this holds at least for
those background events which populate the signal region in phase
space. Such background event will always exist, so any LHC measurement
will always be a statistics exercise. The high energy community has
therefore agreed that we call a five sigma excess over the known
backgrounds a signal:
\begin{alignat}{5}
\frac{S}{\sqrt{B}} &= N_\sigma > 5 \qquad \qquad \qquad 
&&\text{(Gaussian limit)} \notag \\
P_{\rm fluct} &< 5.8 \times 10^{-7} 
&&\text{(fluctuation probability)}
\end{alignat}
Do not trust anybody who wants to sell you a three sigma evidence as
a discovery, even I have seen a great number of those go away. People
often have good personal reasons to advertize such effects, but all
they are really saying is that their errors do not allow them to make
a conclusive statement. This brings us to a well kept secret in the
phenomenology community, which is the important impact of error
bars when we search for exciting new physics. Since for theorists
understanding LHC events and in particular background events means
QCD, we need to understand where our predictions come from and what
they assume, so here we go...

\section{QCD and scales}

Not all processes which involve QCD have to look incredibly
complicated --- let us start with a simple question: we know how to
compute the production rate and distributions for $Z$ production for
example at LEP $e^+e^- \to Z$. To make all phase space integrals
simple, we assume that the $Z$ boson is on-shell, so we can simply add
a decay matrix element and a decay phase space integration for
example compute the process $e^+ e^- \to Z \to \mu^+ \mu^-$.

So here is the question: how do we compute the production of a $Z$
boson at the LHC? This process is usually referred to as
\underline{Drell--Yan production}, even though we will most likely
produce neither Drell nor Yan at the LHC. In our first attempts we
explicitly do not care about additional jets, so if we assume the
proton consists of quarks and gluons we simply compute the process $q
\qb \to Z$ under the assumption that the quarks are partons inside
protons. Modulo the $SU(2)$ and $U(1)$ charges which describe the $Zf
\fb$ coupling
\begin{equation}
 - i \gamma^\mu \left( \ell P_L + r P_R \right)
 \qquad \qquad 
 \ell = \frac{e}{s_w c_w} \; \left( T_3 - Q s_w^2 \right)
 \qquad \quad
 r = \ell \Big|_{T_3=0}
\end{equation}
the matrix element and the squared matrix element for the partonic
process $q \qb \to Z$ will be the same as the corresponding matrix
element squared for $e^+ e^- \to Z$, with an additional color factor.
This color factor counts the number of $SU(3)$ states which can be
combined to form a color singlet like the $Z$. This additional factor
should come out of the color trace which is part of the Feynman rules,
and it is $N_c$. On the other hand, we do not observe color in the
initial state, and the color structure of the incoming $q \qb$ pair
has no impact on the $Z$--production matrix element, so we average
over the color. This gives us another factor $1/N_c^2$ in the
averaged matrix element (modulo factors two)
\begin{equation}
 \MX(q \qb \to Z) \sim
  \frac{1}{4 N_c} \; m_Z^2 \; \left( \ell^2 + r^2 \right) \, .
\end{equation}
Notice that matrix elements we compute from our Feynman rules are not
automatically numbers without a mass unit.  Next, we add the phase
space for a one-particle final state. In four space--time dimensions
(this will become important later) we can compute a total cross
section out of a matrix element squared as
\begin{alignat}{5}
 s \; \frac{d \sigma}{d y} & = \frac{\pi}{(4 \pi)^2} \;
                               \left( 1 - \tau \right) \; \MX \notag \\
\end{alignat}
The mass of the final state appears as $\tau = m_Z^2/s$ and can of
course be $m_W$ or the Higgs mass or the mass of a KK graviton (I know
you smart-asses in the back row!). If we define $s$ as the partonic
invariant mass of the two quarks using the Mandelstam variable $s =
(k_2+k_2)^2 = 2 (k_1 k_2)$, momentum conservation just means $s =
m_Z^2$. This simple one-particle phase space has only one free
parameter, the reduced polar angle $y=(1 + \cos \theta)/2 = 0 \cdots
1$. The azimuthal angle $\phi$ plays no role at colliders, unless you
want to compute gravitational effects on Higgs production at Atlas and
CMS.  Any LHC Monte Carlo will either random-generate a reference
angle $\phi$ for the partonic process or pick one and keep it fixed.
The second option has at least once lead to considerable confusion and
later amusement at the Tevatron, when people noticed that the
behavior of gauge bosons was dominated by gravity, namely gauge
bosons going up or down. So this is not as trivial a statement as you
might think. At this point I remember that every teacher at every
summer schools always feels the need to define their field of
phenomenology --- for example: phenomenologists are theorists who do
useful things and know funny stories about experiment(alist)s.\bigskip

Until now we have computed the same thing as $Z$ production at LEP,
leaving open the question how to describe quarks inside the proton.
For a proper discussion I refer to any good QCD textbook and in
particular the chapter on deep inelastic scattering.  Instead, I will
follow a pedagogical approach which will as fast as possible take us
to the questions we really want to discuss.

If for now we are happy assuming that quarks move collinear with the
surrounding proton, \ie that at the LHC incoming partons have zero
$p_T$, we can simply write a probability distribution for finding a
parton with a certain fraction of the proton's momentum.  For a
momentum fraction $x = 0\cdots 1$ this \underline{parton density
  function} (pdf) is denoted as $f_i(x)$, where $i$ describes the
different partons in the proton, for our purposes $u,d,c,s,g$. All of
these partons we assume to be massless. We can talk about heavy
bottoms in the proton if you ask me about it later.  Note that in
contrast to structure functions a pdf is not an observable, it is
simply a distribution in the mathematical sense, which means it has to
produce reasonably results when integrated over as an integration
kernel. These parton densities have very different behavior --- for
the valence quarks ($uud$) they peak somewhere around $x \lesssim
1/3$, while the gluon pdf is small at $x \sim 1$ and grows very
rapidly towards small $x$. For some typical part of the relevant
parameter space ($x = 10^{-3} \cdots 10^{-1}$) you can roughly think
of it as $f_g(x) \propto x^{-2}$, towards $x$ values it becomes even
steeper. This steep gluon distribution was initially not expected and
means that for small enough $x$ LHC processes will dominantly be gluon
fusion processes.\bigskip

Given the correct definition and normalization of the pdf we can
compute the \underline{hadronic cross section} from its partonic
counterpart as 
\begin{equation}
\sigma_{\rm tot} = \int_0^1 dx_1 \int_0^1 dx_2 \;
                   f_i(x_1) \, f_j(x_2) \;
                   \hat{\sigma}_{ij}(x_1 x_2 S)
\end{equation}
where $i,j$ are the incoming partons with the momentum factions
$x_{i,j}$. The partonic energy of the scattering process is $s=x_1 x_2
S$ with the LHC proton energy $\sqrt{S}=14$~TeV. The partonic cross
section $\hat{\sigma}$ corresponds to the cross sections $\sigma$ we
already discussed. It has to include all the necessary $\Theta$ and
$\delta$ functions for energy--momentum conservation. When we express
a general $n$--particle cross section $\hat{\sigma}$ including the
phase space integration, the $x_i$ integrations and the phase space
integrations can of course be swapped, but Jacobians will make your
life hell when you attempt to get them right. Luckily, there are very
efficient numerical phase space generators on the market which
transform a hadronic $n$--particle phase space integration into a unit
hypercube, so we do not have to worry in our every day life.

\subsection{UV divergences and the renormalization scale} 

Renormalization, \ie the proper treatment of ultraviolet divergences,
is one of the most important aspects of field theories; if you are not
comfortable with it you might want to attend a lecture on field
theory. The one aspect of renormalization I would like to discuss is
the appearance of the renormalization scale. In perturbation theory,
scales arise from the regularization of infrared or ultraviolet
divergences, as we can see writing down a simple loop integral
corresponding to two virtual massive scalars with a momentum $p$
flowing through the diagram:
\begin{equation}
  B(p^2;m,m) \equiv
  \int \frac{d^4q}{16 \pi^2} \; \frac{1}{q^2-m^2} \frac{1}{(q+p)^2-m^2}
\end{equation}
Such diagrams appear for example in the gluon self energy, with
massless scalars for ghosts, with some Dirac trace in the numerator
for quarks, and with massive scalars for supersymmetric scalar quarks.
This integral is UV divergent, so we have to regularize it, express
the divergence in some well-defined manner, and get rid of it by
renormalization. One way is to introduce a cutoff into the momentum
integral $\Lambda$, for example through the so-called Pauli--Villars
regularization. Because the UV behavior of the integrand cannot depend
on IR-relevant parameters, the UV divergence cannot involve the mass
$m$ or the external momentum $p^2$. This means that its divergence has
to be proportional to $\log \Lambda/\mu^2$ with some scale $\mu^2$
which is an artifact of the regularization of such a Feynman diagram.

This question is easier to answer in the more modern
\underline{dimensional regularization}. There, we shift the power of
the momentum integration and use analytic continuation in the number
of space--time dimensions to renormalize the theory
\begin{equation}
  \int \frac{d^4q}{16 \pi^2} \cdots \longrightarrow
  \mu^{2\epsilon} \;  \int \frac{d^{4-2 \epsilon}q}{16 \pi^2} \cdots
  =
  \frac{i \mu^{2\epsilon}}{(4 \pi)^2} \;
                      \left[ \frac{C_{-1}}{\epsilon} 
                           + C_0 
                           + C_1 \, \epsilon + \mathcal{O}(\epsilon^2)
                      \right]
\end{equation}
The constants $C_i$ depend on the loop integral we are considering.
The scale $\mu$ we have to introduce to ensure the matrix element and
the observables, like cross sections, have the usual mass dimensions.
To regularize the UV divergence we pick an $\epsilon>0$, giving us
mathematically well-defined poles $1/\epsilon$. If you compute the
scalar loop integrals you will see that defining them with the
integration measure $1/(i \pi^2)$ will make them come out as of the
order $\mathcal{O}(1)$, in case you ever wondered about factors $1/(4
\pi)^2 = \pi^2/(2 \pi)^4$ which usually end up in front of the loop
integrals.

The poles in $\epsilon$ will cancel with the counter terms, \ie we
renormalize the theory. Counter terms we include by shifting the
renormalized parameter in the leading-order matrix element, \eg
$\MX(g) \to \MX(g+\delta g)$ with a coupling $\delta g \propto
1/\epsilon$, when computing $|\mathcal{M}_{\rm Born} +
\mathcal{M}_{\rm virt}|^2$. If we use a physical renormalization
condition there will not be any free scale $\mu$ in the definition of
$\delta g$. As an example for a physical reference we can think of the
electromagnetic coupling or charge $e$, which is usually defined in
the Thomson limit of vanishing momentum flow through the diagram, \ie
$p^2 \to 0$. What is important about these counter terms is that they
do not come with a factor $\mu^{2 \epsilon}$ in front.\bigskip

So while after renormalization the poles $1/\epsilon$ cancel just
fine, the scale factor $\mu^{2 \epsilon}$ will not be matched between
the UV divergence and the counter term. We can keep track of it by
writing a Taylor series in $\epsilon$ for the prefactor of the
regularized but not yet renormalized integral:
\begin{alignat}{5}
 \mu^{2\epsilon} \; 
  \left[ \frac{C_{-1}}{\epsilon} + C_0 + \mathcal{O}(\epsilon) \right] 
 &= 
 e^{2 \epsilon \log \mu} \;
  \left[ \frac{C_{-1}}{\epsilon} + C_0 + \mathcal{O}(\epsilon) \right] 
  \notag \\ 
 &= \left[ 1 + 2 \epsilon \log \mu + \mathcal{O}(\epsilon^2) \right] \;
  \left[ \frac{C_{-1}}{\epsilon} + C_0 + \mathcal{O}(\epsilon) \right]
  \notag \\ 
 &= \frac{C_{-1}}{\epsilon} 
   + C_0
   + 2 \log \mu \, C_{-1}
   +   \mathcal{O}(\epsilon) 
\label{eq:mu_taylor}
\end{alignat}
We see that the pole $C_{-1}/\epsilon$ gives a finite contribution to the cross
section, involving the \underline{renormalization scale} $\mu_R \equiv
\mu$.\bigskip


Just a side remark for completeness: from eq.(\ref{eq:mu_taylor}) we
see that we should not have just pulled out $\mu^{2 \epsilon}$ out of
the integral, because it leads to a logarithm of a number with a mass
unit.  On the other hand, from the way we split the original integral
we know that the remaining $(4-2\epsilon)$-dimensional integral has to
includes logarithms of the kind $\log m^2$ or $\log p^2$ which
re-combine with the $\log \mu^2$ for example to a properly defined
$\log \mu/m$. The only loop integral which has no intrinsic mass scale
is the two-point function with zero mass in the loop and zero momentum
flowing through the integral: $B(p^2=0;0,0)$. It appears for example
as a self-energy correction of external quarks and gluons.  Based on
these dimensional arguments this integral has to be zero, but with a
subtle cancellation of the UV and the IR divergences which we can
schematically write as $1/\epsilon_{\rm IR} - 1/\epsilon_{\rm
  UV}$. Actually, I am thinking right now if following this argument
this integral has to be zero or if it can still be a number, like
2376123/67523, but it definitely has to be finite... And it is zero if
you compute it. \bigskip

Instead of discussing different renormalization schemes and their
scale dependences, let us instead compute a simple renormalization
scale dependent parameter, namely the \underline{running strong
  coupling} $\alpha_s(\mu_R)$. It does not appear in our Drell--Yan
process at leading order, but it does not hurt to know how it appears
in QCD calculations. The simplest process we can look at is two-jet
production at the LHC, where we remember that in some energy range we
will be gluon dominated: $gg \to q \qb$.  The Feynman diagrams include
an $s$--channel off-shell gluon with a momentum flow $p^2 \equiv
s$. At next-to-leading order, this gluon propagator will be corrected
by self-energy loops, where the gluon splits into two quarks or gluons
and re-combines before it produces the two final-state partons.

The gluon self energy correction (or vacuum polarization, as
propagator corrections to gauge bosons are often labelled) will be a
scalar, \ie fermion loops will be closed and the Dirac trace is closed
inside the loop. In color space the self energy will (hopefully) be
diagonal, just like the gluon propagator itself, so we can ignore the
color indices for now. In Minkowski space the gluon propagator in
unitary gauge is proportional to the transverse tensor $T^{\mu \nu} =
g^{\mu\nu} - p^\nu p^\mu/p^2$. The same is true for the gluon self
energy, which we write as $\Pi^{\mu \nu} \equiv \Pi \, T^{\mu \nu}$.
The one useful thing to remember is the simple relation $T^{\mu \nu}
T_\nu^\rho = T^{\mu \rho}$ and $T^{\mu \nu} g_\nu^\rho = T^{\mu
\rho}$. Including the gluon, quark, and ghost loops the regularized 
gluon self energy with a momentum flow $p^2$ reads
\begin{alignat}{5}
 \frac{1}{p^2} \; \Pi\left( \frac{\mu_R^2}{p^2} \right) 
    &=&&     \frac{\alpha_s}{4 \pi} \;
             \left( - \frac{1}{\epsilon} 
                    - \log \frac{\mu_R^2}{p^2} \right) \;
             \left( \frac{13}{6} N_c - \frac{2}{3} n_f \right) 
             + \mathcal{O} (\log m_t^2 ) \notag \\
    &\longrightarrow \;&& 
            \frac{\alpha_s}{4 \pi} \;
             \left( - \frac{1}{\epsilon} 
                    - \log \frac{\mu_R^2}{p^2} \right) \;
             \beta_g
             + \mathcal{O} (\log m_t^2 ) \notag \\
   &&&\text{with} \qquad
  \beta_g = \frac{11}{3} N_c - \frac{2}{3} n_f \, .
\end{alignat}
In the second step we have sneaked in additional contributions to the
renormalization of the strong coupling from the other one-loop
diagrams in the process. The number of fermions coupling to the gluons
is $n_f$. We neglect the additional terms $\log (4\pi)$ and $\log
\gamma_E$ which come with the poles in dimensional regularization.
From the comments on the function $B(p^2;0,0)$ before we could have
guessed that the loop integrals will only give a logarithm $\log p^2$
which then combines with the scale logarithm $\log \mu_R^2$. The
finite top mass actually leads to an additional logarithms which we
omit for now --- this zero-mass limit of our field theory is actually
special and referred to as its conformal limit.

Lacking a well-enough motivated reference point (in the Thomson limit
the strong coupling is divergent, which means QCD is confined towards
large distances and asymptotically free at small distances) we are
tempted to renormalize $\alpha_s$ by also absorbing the scale into the
counter term, which is called the $\overline{\rm MS}$ scheme. It gives
us a running coupling $\alpha_s(p)$. In other words, for a given
momentum transfer $p^2$ we cancel the UV pole and at the same time
shift the strong coupling, after including all relative ($-$) signs, by
\begin{equation}
 \alpha_s \longrightarrow 
 \alpha_s(\mu_R^2) \; \left( 1 
               - \frac{1}{p^2} \; \Pi \left( \frac{\mu_R^2}{p^2} \right)
             \right)
 = 
 \alpha_s(\mu_R^2) \; \left( 1
               - \frac{\alpha_s}{4 \pi} \;
                 \beta_g \;
                 \log \frac{p^2}{\mu_R^2} 
             \right) \, .
\label{eq:shift_alpha}
\end{equation}
\bigskip

We can do even better: the problem with the correction to $\alpha_s$
is that while it is perturbatively suppressed by the usual factor
$\alpha_s/(4 \pi)$ it includes a logarithm which does not need to be
small. Instead of simply including these gluon self-energy
corrections at a given order in perturbation theory we can instead
include all chains with $\Pi$ appearing many times in the off-shell
gluon propagator.  Such a series means we replace the off-shell gluon
propagator by (schematically written)
\begin{alignat}{5}
\frac{T^{\mu \nu}}{p^2}
\longrightarrow &
  \frac{T^{\mu \nu}}{p^2}
 + \left ( \frac{T}{p^2} \cdot (- T \, \Pi) \cdot \frac{T}{p^2} 
    \right)^{\mu \nu} \notag \\
 & \hspace*{20pt}
 + \left( \frac{T}{p^2} \cdot (- T \, \Pi) \cdot \frac{T}{p^2} 
                        \cdot (- T \, \Pi) \cdot \frac{T}{p^2} 
   \right)^{\mu \nu} + \cdots \notag \\
= & \;
  \frac{T^{\mu \nu}}{p^2} \;
  \sum_{j=0}^\infty \left( - \frac{\Pi}{p^2} \right)^j
= \frac{T^{\mu \nu}}{p^2} \; \frac{1}{1 + \Pi/p^2}
\end{alignat}
To avoid indices we abbreviate $T^{\mu \nu} T_\nu^\rho = T \cdot T$
which can be simplified using $(T \cdot T \cdot T)^{\mu \nu} = T^{\mu
\rho} T^\sigma_\rho T_\sigma^\nu = T^{\mu \nu}$.  This re-summation of
the logarithm which occurs in the next-to-leading order corrections to
$\alpha_s$ moves the finite shift in $\alpha_s$ shown in
eq.(\ref{eq:shift_alpha}) into the denominator:
\begin{equation}
 \alpha_s \longrightarrow 
 \alpha_s(\mu_R^2) \; \left( 1
               + \frac{\alpha_s}{4 \pi} \;
                 \beta_g \;
                 \log \frac{p^2}{\mu_R^2} 
             \right)^{-1}
\end{equation}
If we interpret the renormalization scale $\mu_R$ as one reference
point $p_0$ and $p$ as another, we can relate the values of $\alpha_s$
between two reference points with a \underline{renormalization group
  equation} (RGE) which evolves physical parameters from one scale to
another:
\begin{alignat}{5}
 \alpha_s(p^2)&= \alpha_s(p_0^2) \; 
                 \left( 1
               + \frac{\alpha_s(p_0^2)}{4 \pi} \;
                 \beta_g \;
                 \log \frac{p^2}{p_0^2} 
                 \right)^{-1} \notag \\
 \frac{1}{\alpha_s(p^2)} &=
                 \frac{1}{\alpha_s(p_0^2)}
                 \left( 1
               + \frac{\alpha_s(p_0^2)}{4 \pi} \;
                 \beta_g \;
                 \log \frac{p^2}{p_0^2} 
                 \right)
               = \frac{1}{\alpha_s(p_0^2)}
               + \frac{1}{4\pi} \; \beta_g \;
                 \log \frac{p^2}{p_0^2}
\end{alignat}
The factor $\alpha_s$ inside the parentheses can be evaluated at any
of the two scales, the difference is going to be a higher-order
effect.  The interpretation of $\beta_g$ is now obvious: when we
differentiate the shifted $\alpha_s(p^2)$ with respect to the
momentum transfer $p^2$ we find:
\begin{equation}
 \frac{1}{\alpha_s} \;
 \frac{d \alpha_s}{d \log p^2} = - \frac{\alpha_s}{4 \pi} \beta_g
 \qquad \qquad \text{or} \qquad \qquad
 \frac{1}{g_s} \; 
 \frac{d g_s}{d \log p} 
                                 = - \frac{\alpha_s}{4 \pi} \beta_g
                                 = - g_s^2 \beta_g
\end{equation}
This is the famous running of the strong coupling constant! \bigskip

Before we move on, let us collect the logic of the argument given in
this section: when we regularize an UV divergence we automatically
introduce a reference scale. Naively, this could be a UV cutoff scale,
but even the seemingly scale invariant dimensional regularization
cannot avoid the introduction of a scale, even in the conformal limit
of our theory. There are several ways of dealing with such a scale:
first, we can renormalize our parameter at a reference point.
Secondly, we can define a running parameter, \ie absorb the scale
logarithm into the $\overline{\rm MS}$ counter term. This way, at each
order in perturbation theory we can translate values for example of
the strong coupling from one momentum scale to another momentum scale.
If we are lucky, we can re-sum these logarithms to all orders in
perturbation theory, which gives us more precise perturbative
predictions even in the presence of large logarithms, \ie large scale
differences for our renormalized parameters. Such a (re--) summation
is linked with the definition of scale dependent parameters.

\subsection{IR divergences and the factorization scale} 

After this brief excursion into renormalization and UV divergences we
can return to the original example, the Drell--Yan process at the LHC.
In our last attempt we wrote down the hadronic cross sections in terms
of parton distributions at leading order. These pdfs are only
functions of the (collinear) momentum fraction of the partons in the
proton.

The perturbative question we need to ask for this process is: what
happens if we radiate additional jets which for one reason or another
we do not observe in the detector. Throughout this writeup I will use
the terms \underline{jets and final state partons} synonymously, which
is not really correct once we include jet algorithms and
hadronization. On the other hand, in most cases a jet algorithms is
designed to take us from some kind of energy deposition in the
calorimeter to the parton radiated in the hard process. This is
particularly true for modern developments like the so-called matrix
element method to measure the top mass. Recently, people have looked
into the question what kind of jets come from very fast collimated $W$
or top decays and how such fat jets could be identified looking into
the details of the jet algorithm. But let's face it, you can try to do
such analyses after you really understand the QCD of hard processes,
and you should not trust such analyses unless they come from groups
which know a whole lot of QCD and preferable involve experimentalists
who know their calorimeters very well.\bigskip

So let us get back to the radiation of additional partons in the
Drell--Yan process. These can for example be gluons radiated from the
incoming quarks. This means we can start by compute the cross section
for the partonic process $q \qb \to Z g$. However, this partonic
process involves renormalization as well as an avalanche of loop
diagrams which have to be included before we can say anything
reasonable, \ie UV and IR finite. Instead, we can look at the crossed
process $q g \to Z q$, which should behave similarly as a $2 \to 2$
process, except that it has a different incoming state than the
leading-order Drell--Yan process and hence no virtual
corrections. This means we do not have to deal with renormalization
and UV divergences and can concentrate on parton or jet radiation from
the initial state.\bigskip

The amplitude for this $2 \to 2$ process is --- modulo the charges and
averaging factors, but including all Mandelstam variables
\begin{equation}
 \MX  \propto 8 \left[- \frac{t}{s}
                      - \frac{s}{t}
                      + \frac{2 m_Z^2 (s + t - m_Z^2)}{st}
                \right]
\end{equation}
The new Mandelstam variables can be expressed in terms of the rescaled
gluon-emission angle $y=(1 + \cos \theta)/2$ as $t = -s (1-\tau) y$
and $u = -s (1-\tau) (1-y)$. As a sanity check we can confirm that
$t+u=-s+m_Z^2$. The collinear limit when the gluon is radiated in the
beam direction is given by $y \to 0$, which corresponds to $t
\to 0$ with finite $u=-s+m_Z^2$. In that case the matrix element
becomes
\begin{equation}
 \MX \sim 8 \left[ \frac{s^2 - 2 s m_Z^2 + 2 m_Z^4}{s(s-m_Z^2)} \; \frac{1}{y}
                 - \frac{2 m_Z^2}{s} \;
                 + \mathcal{O}(y)
            \right]
\end{equation}
This expression is divergent for collinear gluon radiation, \ie for
small angles $y$. We can translate this $1/y$ divergence for example
into the transverse momentum of the gluon or $Z$ according to
\begin{equation}
s p_T^2 
= t u 
= s^2 (1 - \tau)^2 \; y (1-y) 
= (s-m_Z^2)^2 y + \mathcal{O}(y^2)
\end{equation}
In the collinear limit our matrix element squared then becomes
\begin{equation}
 \MX \sim 8 \left[ \frac{s^2 - 2 s m_Z^2 + 2 m_Z^4}{s^2} \;
                   \frac{(s-m_Z^2)}{p_T^2}
                 + \mathcal{O}(p_T^0)
            \right] \, .
\end{equation}
The matrix element for the tree-level process $q g \to Z q$ diverges
like $1/p_T^2$. To compute the total cross section for this process we
need to integrate it over the two-particle phase space.  Without deriving
this result we quote that this integration can be written in the
transverse momentum of the outgoing particles, in which case the
Jacobian for this integration introduces a factor $p_T$.
Approximating the matrix element as $C/p_T^2$, we have to integrate
\begin{alignat}{5}
  \int_{y^{\rm min}}^{y^{\rm max}} d y \frac{C}{y}
= \int_{p_T^{\rm min}}^{p_T^{\rm max}} d p_T^2 \frac{C}{p_T^2}
=& \; 2 \int_{p_T^{\rm min}}^{p_T^{\rm max}} d p_T \; p_T \; \frac{C}{p_T^2}
\notag \\
\simeq& \; 2 C \int_{p_T^{\rm min}}^{p_T^{\rm max}} d p_T \frac{1}{p_T}
= 2 C \; \log \frac{p_T^{\rm max}}{p_T^{\rm min}}
\label{eq:collinear}
\end{alignat}
The form $C/p_T^2$ for the matrix element is of course only valid in
the collinear limit; in the remaining phase space $C$ is not a
constant.  However, this formula describes well the collinear IR
divergence arising from gluon radiation at the LHC (or photon
radiation at $e^+ e^-$ colliders, for that matter).\bigskip

We can follow the same strategy as for the UV divergence.  First, we
regularize the divergence using dimensional regularization, and then
we find a well-defined way to get rid of it. Dimensional
regularization now means we have to write the two-particle phase space
in $n=4-2 \epsilon$ dimensions. Just for the fun, here is the complete
formula in terms of $y$:
\begin{equation}
 s \; \frac{d \sigma}{d y} =
  \frac{\pi (4 \pi)^{-2+\epsilon}}{\Gamma(1-\epsilon)} \;
  \left( \frac{\mu^2}{m_Z^2} \right)^\epsilon \;
  \frac{\tau^\epsilon (1-\tau)^{1-2 \epsilon}}
       {y^\epsilon (1-y)^\epsilon} \MX
 \sim 
  \left( \frac{\mu^2}{m_Z^2} \right)^\epsilon \;
  \frac{\MX}{y^\epsilon (1-y)^\epsilon} \, .
\label{eq:phase}
\end{equation}
In the second step we only keep the factors we are interested in. The
additional factor $y^{-\epsilon}$ regularizes the integral at $y \to
0$, as long as $\epsilon<0$, which just slightly increases the
suppression of the integrand in the IR regime. After integrating the
leading term $1/y^{1+\epsilon}$ we have a pole
$1/(-\epsilon)$. Obviously, this regularization procedure is symmetric
in $y \leftrightarrow (1-y)$. What is important to notice is again the
appearance of a scale $\mu^{2 \epsilon}$ with the $n$-dimensional
integral. This scale arises from the IR regularization of the phase
space integral and is referred to as \underline{factorization scale}
$\mu_F$.\bigskip

From our argument we can safely guess that the same divergence which
we encounter for the process $q g \to Z q$ will also appear in the
crossed process $q \qb \to Z g$, after cancelling additional soft IR
divergences between virtual and real gluon emission diagrams. We can
write all these collinear divergences in a universal form, which is
independent of the hard process (like Drell--Yan production). In the
collinear limit, the probabilities of radiating additional partons or
splitting into additional partons is given by universal
\underline{splitting functions}, which govern the collinear behavior
of the parton-radiation cross section:
\begin{equation}
  \frac{1}{\sigma_{\rm tot}} \; d \sigma
  \sim 
  \frac{\alpha_s}{2 \pi} \; \frac{dy}{y} \; dx \; P_j(x)
  =
  \frac{\alpha_s}{2 \pi} \; \frac{dp_T^2}{p_T^2} \; dx \; P_j(x)
\end{equation}
The momentum fraction which the incoming parton transfers to the
parton entering the hard process is given by $x$. The rescaled angle
$y$ is one way to integrate over the transverse-momentum space. The
splitting kernels are different for different partons involved:
\begin{alignat}{5}
P_{q \leftarrow q}(x) &= C_F \; \frac{1+x^2}{1-x} \qquad \qquad \qquad
P_{g \leftarrow q}(x)  = C_F \; \frac{1+(1-x)^2}{x} \notag \\
P_{q \leftarrow g}(x) &= T_R \; \left( x^2 + (1-x)^2 \right) \notag \\
P_{g \leftarrow g}(x) &= C_A \; \left(  \frac{x}{1-x}
                                      + \frac{1-x}{x}
                                      + x (1-x) 
                               \right)
\label{eq:splitting}
\end{alignat}
The underlying QCD vertices in these four collinear splittings are the
$qqg$ and $ggg$ vertices. This means that a gluon can split
independently into a pair of quarks and a pair of gluons. A quark can
only radiate a gluon, which implies $P_{q \leftarrow q}(1-x) = P_{g
\leftarrow q}(x)$, depending on which of the two final state partons
we are interested in. For these formulas we have sneaked in the
Casimir factors of $SU(N)$, which allow us to generalize our approach
beyond QCD. For practical purposes we can insert the SU(3) values
$C_F= (N_c^2-1)/(2 N_c) = 4/3$, $C_A = N_c = 3$ and $T_R = 1/2$. Once
more looking at the different splitting kernels we see that in the
soft-daughter limit $x \to 0$ the daughter quarks $P_{q \leftarrow
q}$ and $P_{q \leftarrow g}$ are well defined, while the gluon
daughters $P_{g \leftarrow q}$ and $P_{g \leftarrow g}$ are infrared
divergent.\bigskip

What we need for our partonic subprocess $q g \to Z q$ is the
splitting of a gluon into two quarks, one of which then enters the
hard Drell--Yan process. In the \underline{collinear limit} this
splitting is described by $P_{q \leftarrow g}$. We explicitly see that
there is no additional soft singularity for vanishing quark energy,
only the collinear singularity in $y$ or $p_T$. This is good news,
since in the absence of virtual corrections we would have no idea how
to get rid of or cancel this soft divergence.\bigskip

\begin{figure}[t]
\begin{center}
\includegraphics[width=4.5cm,angle=0]{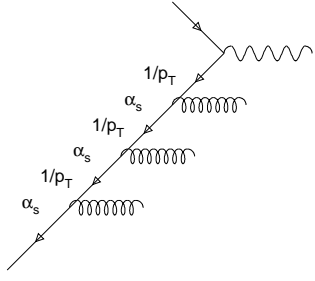}
\end{center}
\vspace*{-0.5cm}
\caption{Feynman diagrams for the repeated emission of a gluon 
 from the incoming leg of a Drell--Yan process. The labels indicate
 the appearance of $\alpha_s$ as well as the leading divergence of the
 phase space integration.}
\label{fig:coll}
\end{figure}

If we for example consider repeated collinear gluon emission off an
incoming quark leg, we naively get a correction suppressed by powers
of $\alpha_s$, because of the strong coupling of the gluon. Such a
chain of gluon emissions is illustrated in Fig.~\ref{fig:coll}. On the
other hand, the $y$ integration over each new final state gluon
combined with the $1/y$ or $1/p_T$ divergence in the matrix element
squared leads to a possibly large logarithm which can be easiest
written in terms of the upper and lower boundary of the $p_T$
integration. This means, at higher orders we expect corrections of the
form
\begin{equation}
 \sigma_{\rm tot} \sim \sum_j C_j \; 
                       \left( \alpha_s 
                           \log \frac{p_T^{\rm max}}{p_T^{\rm min}}
                       \right)^j
\end{equation}
with some factors $C_j$. Because the splitting probability is
universal, these fixed-order corrections can be re-summed to all
orders, just like the gluon self energy. You notice how successful
perturbation theory becomes every time we encounter a geometric
series? And again, in complete analogy with the gluon self energy,
this universal factor can be absorbed into another quantity, which are
the parton densities.\bigskip

However, there are three important differences to the running
coupling:

First, we are now absorbing IR divergences into running parton
densities. We are not renormalizing them, because renormalization is a
well-defined procedure to absorb UV divergences into a redefined
Lagrangian.

Secondly, the quarks and gluons split into each other, which means
that the parton densities will form a set of coupled differential
equations which describe their running instead of a simple
differential equation with a beta function.

And third, the splitting kernels are not just functions to multiply
the parton densities, but they are integration kernels, so we end up
with a coupled set of integro-differential equations which describe
the parton densities as a function of the factorization scale. These
equation are called the Dokshitzer--Gribov--Lipatov--Altarelli--Parisi
or \underline{DGLAP equations}
\begin{equation}
  \frac{d f_i(x,\mu_F)}{d \log \mu_F^2} 
  = \frac{\alpha_s}{2 \pi} \;
    \sum_j \; \int_x^1 \; \frac{dx'}{x'} \;
    P_{i \leftarrow j}\left( \frac{x}{x'} \right) \;
    f_j(x',\mu_F) \, .
\label{eq:dglap}
\end{equation}
We can discuss this formula briefly: to compute the scale dependence
of a parton density $f_i$ we have to consider all partons $j$ which
can split into $i$. For each splitting process, we have to integrate
over all momentum fractions $x'$ which can lead to a momentum fraction
$x$ after splitting, which means we have to integrate $z$ from $x$ to
1. The relative momentum fraction in the splitting is then $x/z < 1$.

The DGLAP equation by construction resums collinear logarithms. There
is another class of logarithms which can potentially become large,
namely soft logarithms $\log x$, corresponding to the soft divergence
of the diagonal splitting kernels. This reflects the fact that if you
have for example a charged particle propagating there are two ways to
radiate photons without any cost in probability, either collinear
photons or soft photons. We know from QED that both of these effects
lead to finite survival probabilities once we sum up these collinear
and soft logarithms. Unfortunately, or fortunately, we have not seen
any experimental evidence of these soft logarithms dominating the
parton densities yet, so we can for now stick to DGLAP.

Going back to our original problem, we can now write the hadronic
cross section production for Drell--Yan production or other LHC
processes as:
\begin{equation}
\sigma_{\rm tot}(\mu_F,\mu_R) 
             = \int_0^1 dx_1 \int_0^1 dx_2 \;
               f_i(x_1,\mu_F) \, f_j(x_2, \mu_F) \;
               \hat{\sigma}_{ij}(x_1 x_2 S, \mu_R)
\label{eq:hadronic}
\end{equation}
Since our particular Drell--Yan process at leading order only involves
weak couplings, it does not include $\alpha_s$ at leading order.  We
will only see $\alpha_s$ and with it a renormalization scale $\mu_R$
appear at next-to-leading order, when we include an additional
final state parton.\bigskip

After this derivation, we can attempt a \underline{physical
interpretation} of the factorization scale. The collinear divergence
we encounter for example in the $q g \to Z q$ process is absorbed into
the parton densities using the universal collinear splitting kernels.
In other words, as long as the $p_T$ distribution of the matrix
element follows eq.(\ref{eq:collinear}), the radiation of any number
of additional partons from the incoming partons is now included.
These additional partons or jets we obviously cannot veto without
getting into perturbative hell with QCD.  This is why we should really
write $pp \to Z + X$ when talking about factorization-scale dependent
parton densities as defined in eq.(\ref{eq:hadronic}).\bigskip

If we look at the $d \sigma/d p_T$ distribution of additional partons
we can divide the entire phase space into two regions. The collinear
region is defined by the leading $1/p_T$ behavior.  At some point the
$p_T$ distribution will then start decreasing faster, for example
because of phase space limitations.  The transition scale should
roughly be the factorization scale.  In the DGLAP evolution we
approximate all parton radiation as being collinear with the hadron,
\ie move them from the region $p_T < \mu_F$ onto the point $p_T =
0$. This kind of $p_T$ spectrum can be nicely studied using bottom
parton densities. They have the advantage that there is no intrinsic
bottom content in the proton. Instead, all bottoms have to arise from
gluon splitting, which we can compute using perturbative QCD. If we
actually compute the bottom parton densities, the factorization scale
is not an unphysical free parameter, but it should at least roughly
come out of the calculation of the bottom parton densities. So we can
for example compute the bottom-induced process $b\bar{b} \to H$
including resummed collinear logarithms using bottom densities or
derive it from the fixed-order process $gg \to b\bar{b}H$. When
comparing the $p_{T,b}$ spectra it turns out that the bottom
factorization scale is indeed proportional to the Higgs mass (or hard
scale), but including a relative factor of the order $1/4$. If we
naively use $\mu_F = m_H$ we will create an inconsistency in the
definition of the bottom parton densities which leads to large
higher-order corrections.

Going back to the $p_T$ spectrum of radiated partons or jets --- when
the transverse momentum of an additional parton becomes large enough
that the matrix element does not behave like eq.(\ref{eq:collinear})
anymore, this parton is not well described by the collinear parton
densities. We should definitely choose $\mu_F$ such that this
high-$p_T$ range is not governed by the DGLAP equation. We actually
have to compute the hard and now finite matrix elements for $pp \to
Z+$jets to predict the behavior of these jets. How to combine
collinear jets as they are included in the parton densities and hard
partonic jets is what the rest of this lecture will be about.

\begin{table}[b]
\begin{small}
\begin{tabular}{l||l|l}
   & renormalization scale $\mu_R$
   & factorization scale $\mu_F$ \\ \hline
 source
   & ultraviolet divergence
   & collinear (infrared) divergence \\[2mm]
 poles cancelled
   & counter terms 
   & parton densities \\
   & (renormalization)
   & (mass factorization) \\
 summation
   & resum self energy bubbles
   & resum collinear logarithms \\
 parameter 
   & running coupling $\alpha_s(\mu_R)$
   & parton density $f_j(x, \mu_F)$ \\
 evolution 
   & RGE for $\alpha_s$
   & DGLAP equation \\[2mm]
 large scales
   & typically decrease of $\sigma_{\rm tot}$
   & typically increase of $\sigma_{\rm tot}$ \\[2mm]
 theory
   & renormalizability
   & factorization \\
   & proven for gauge theories
   & proven all order for DIS \\
   && proven order-by-order DY... \\
\end{tabular}
\end{small}
\caption{Comparison of renormalization and factorization scales 
  appearing in LHC cross sections.}
\label{tab:scales}
\end{table}

\subsection{Right or wrong scales}

Looking back at the last two sections we introduce the factorization
and renormalization scales completely in parallel.  First, computing
perturbative higher-order contributions to scattering amplitudes we
encounter divergences. Both of them we regularize, for example using
dimensional regularization (remember that we had to choose $n=4 - 2
\epsilon<4$ for UV and $n>4$ for IR divergences). After absorbing the
divergences into a re-definition of the respective parameters,
referred to as renormalization for example of the strong coupling in
the case of an UV divergence and as mass factorization absorbing IR
divergences into the parton distributions we are left with a scale
artifact. In both cases, this redefinition was not perturbative at
fixed order, but involved summing possibly large logarithms.  The
evolution of these parameters from one renormalization/factorization
scale to another is described either by a simple beta function in the
case of renormalization and by the DGLAP equation in the case of mass
factorization. There is one formal difference between these two
otherwise very similar approaches. The fact that we can actually
absorb UV divergences into process-independent universal counter terms
is called renormalizability and has been proven to all orders for the
kind of gauge theories we are dealing with. The universality of IR
splitting kernels has not (yet) in general been proven, but on the
other hand we have never seen an example where is failed. Actually,
for a while we thought there might be a problem with factorization in
supersymmetric theories using the supersymmetric version of the
$\overline{\rm MS}$ scheme, but this has since been resolved. A
comparison of the two relevant scales for LHC physics is shown in
Tab.~\ref{tab:scales}\bigskip

The way I introduced factorization and renormalization scales clearly
describes an artifact of perturbation theory and the way we have to
treat divergences. What actually happens if we include all orders in
perturbation theory? In that case for example the resummation of the
self-energy bubbles is simply one class of diagrams which have to be
included, either order-by-order or rearranged into a resummation. For
example the two jet production rate will then not depend on
arbitrarily chosen renormalization or factorization scales
$\mu$. Within the expression for the cross section, though, we know
from the arguments above that we have to evaluate renormalized
parameters at some scale. This scale dependence will cancel once we
put together all its implicit and explicit appearances contributing to
the total rate at all orders. In other words, whatever scale we
evaluate the strong couplings at gets compensated by other scale
logarithms in the complete expression. In the ideal case, these
logarithms are small and do not spoil perturbation theory by inducing
large logarithms. If we think of a process with one distinct external
scale, like the $Z$ mass, we know that all these logarithms have the
form $\log \mu/m_Z$. This logarithm is truly an artifact, because it
would not need to appear if we evaluated everything at the `correct'
external energy scale of the process, namely $m_Z$. In that sense we
can even think of the running coupling as an \underline{running
  observable}, which depends on the external energy of the
process. This energy scale is not a perturbative artifact, but the
cross section even to all orders really depends on the external energy
scale. The only problem is that most processes after analysis cuts
have more than one scale.\bigskip

We can turn this argument around and estimate the
minimum \underline{theory error} on a prediction of a cross section to
be given by the scale dependence in an interval around what we would
consider a reasonable scale. Notice that this error estimate is not at
all conservative; for example the renormalization scale dependence of
the Drell--Yan production rate is zero, because $\alpha_s$ only enters
are next-to-leading order. At the same time we know that the
next-to-leading order correction to the cross section at the LHC is of
the order of 30\%, which far exceeds the factorization scale
dependence.\bigskip

Guessing the right scale choice for a process is also hard. For
example in leading-order Drell--Yan production there is one scale,
$m_Z$, so any scale logarithm (as described above) has to be $\log
\mu/m_Z$. If we set $\mu = m_Z$ all scale logarithms will vanish. In
reality, any observable at the LHC will include several different
scales, which do not allow us to just define just one `correct'
scale. On the other hand, there are definitely completely wrong scale
choices. For example, using $1000 \times m_Z$ as a typical scale in
the Drell--Yan process will if nothing else lead to logarithms of the
size $\log 1000$ whenever a scale logarithm appears. These logarithms
have to be cancelled to all orders in perturbation theory, introducing
unreasonably large higher-order corrections.

When describing jet radiation, people usually introduce a phase-space
dependent renormalization scale, evaluating $\alpha_s(p_{T,j})$. This
choice gives the best kinematic distributions for the additional
partons, but to compute a cross section it is the one scale choice
which is forbidden by QCD and factorization: scales can only depend on
exclusive observables, \ie momenta which are given after integrating
over the phase space. For the Drell--Yan process such a scale could be
$m_Z$, or the mass of heavy new-physics states in their production
process. Otherwise we double-count logarithms and spoil the collinear
resummation. But as long as we are mostly concerned with
distributions, we even use the transverse-momentum scale very
successfully. To summarize this brief mess: while there is no such
thing as the correct scale choice, there are more or less smart
choices, and there are definitely very wrong choices, which lead to an
unstable perturbative behavior.\bigskip

Of course, these sections on divergences and scales cannot do the
topic justice. They fall short left and right, hardly any of the
factors are correct (they are not that important either), and I am
omitting any formal derivation of this resummation technique for the
parton densities. On the other hand, we can derive some general
message from them: because we compute cross sections in perturbation
theory, the absorption of ubiquitous UV and IR divergences
automatically lead to the appearance of scales. These scales are
actually useful because running parameters allow us to resum
logarithms in perturbation theory, or in other words allow us to
compute certain dominant effects to all orders in perturbation theory,
in spite of only computing the hard processes at a given loop order.
This means that any LHC observable we compute will depend on the
factorization and renormalization scales, and we have to learn how to
either get rid of the scale dependence by having the Germans compute
higher and higher loop orders, or use the Californian/Italian approach
to derive useful scale choices in a relaxed atmosphere, to make use of
the resummed precision of our calculation.

\section{Hard vs collinear jets}

Jets are a major problem we are facing at the Tevatron and will be the
most dangerous problem at the LHC. Let's face it, the LHC is not built
do study QCD effects. To the contrary, if we wanted to study QCD, the
Tevatron with its lower luminosity would be the better place to do so.
Jets at the LHC by themselves are not interesting, they are a nuisance
and they are the most serious threat to the success of the LHC
program.

The main difference between QCD at the Tevatron and QCD at the LHC is
the energy scale of the jets we encounter.  \underline{Collinear jets}
or jets with a small transverse momentum, are well described by
partons in the collinear approximation and simulated by a
\underline{parton shower}.  This parton shower is the attempt to undo
the approximation $p_T \to 0$ we need to make when we absorb collinear
radiation in parton distributions using the DGLAP equation. Strictly
speaking, the parton shower can and should only fill the phase space
region $p_T = 0...\mu_F$ which is not covered by explicit additional
parton radiation. Such so-called
\underline{hard jets} or jets with a large transverse momentum are
described by hard matrix elements which we can compute using the QCD
Feynman rules. Because of the logarithmic enhancement we have observed
for collinear additional partons, there are much more collinear and
soft jets than hard jets.\bigskip

The problem at the LHC is the range of `soft' or `collinear' and
`hard'. As mentioned above, we can define these terms by the validity
of the collinear approximation in eq.(\ref{eq:collinear}). The maximum
$p_T$ of a collinear jet is the region for which the jet radiation
cross section behaves like $1/p_T$. We know that for harder and harder
jets we will at some point become limited by the partonic energy
available at the LHC, which means the $p_T$ distribution of additional
jets will start dropping faster than $1/p_T$. At this point the
logarithmic enhancement will cease to exist, and jets will be
described by the regular matrix element squared without any
resummation.

Quarks and gluons produced in association with gauge bosons at the
Tevatron behave like collinear jets for $p_T \lesssim 20$~GeV, because
the quarks at the Tevatron are limited in energy. At the LHC, jets
produced in association with tops behave like collinear jets to $p_T
\sim 150$~GeV, jets produced with 500~GeV gluinos behave like
collinear jets to $p_T$ scales larger than 300~GeV. This is not good
news, because collinear jets means many jets, and many jets produce
\underline{combinatorical backgrounds} or ruin the missing momentum
resolution of the detector. Maybe I should sketch the notion of
combinatorical backgrounds: if you are looking for example for two
jets to reconstruct an invariant mass you can simply plot all events
as a function of this invariant mass and cut the background by
requiring all event to sit around a peak in $m_{jj}$. However, if you
have for example three jets in the event you have to decide which of
the three jet-jet combinations should go into this distribution. If
this seems not possible, you can alternatively consider two of the
three combinations as uncorrelated `background' events. In other
words, you make three histogram entries out of your signal or
background event and consider all background events plus two of the
three signal combinations as background. This way the
signal-to-background ratio decreases from $N_S/N_B$ to
$N_S/(3N_B+2N_S)$, \ie by at least a factor of three. You can guess
that picking two particles out of four candidates with its six
combinations has great potential to make your analysis a candidate for
this circular folder under your desk. The most famous victim of such
combinatorics might be the formerly promising Higgs discovery channel
$pp \to t\bar{t}H$ with $H \to b\bar{b}$.\bigskip

All this means for theorists that at the LHC we have to learn how to
model collinear and hard jets reliably. This is what the remainder of
the QCD lectures will be about. Achieving this understanding I
consider the most important development in QCD since I started working
on physics.  Discussing the different approaches we will see why such
general--$p_T$ jets are hard to understand and even harder to properly
simulate.

\subsection{Sudakov factors}

Before we discuss any physics it makes sense to introduce the
so-called Sudakov factors which will appear in the next sections. This
technical term is used by QCD experts to ensure that other LHC
physicists feel inferior and do not get on their nerves. But, really,
Sudakov factors are nothing but simple survival probabilities. Let us
start with an event which we would expect to occur $p$ times, given
its probability and given the number of shots. The probability of
observing it $n$ times is given by the Poisson distribution
\begin{equation}
 \mathcal{P}(n;p) = \frac{p^n \, e^{-p}}{n!} \, .
\end{equation}
This distribution will develop a mean at $p$, which means most of the
time we will indeed see about the expected number of events. For large
numbers it will become a Gaussian. In the opposite direction, using
this distribution we can compute the probability of observing zero
events, which is $\mathcal{P}(0;p) = e^{-p}$.  This formula comes in
handy when we want to know how likely it is that we do not see a
parton splitting in a certain energy range.\bigskip

According to the last section, the differential probability of a
parton to split or emit another parton at a scale $\mu$ and with the
daughter's momentum fraction $x$ is given by the splitting kernel
$P_{i \leftarrow j}(x)$ times $d p_T^2/p_T^2$. This energy measure is
a little tricky because we compute the splitting kernels in the
collinear approximation, so $p_T^2$ is the most inconvenient
observable to use. We can approximately replace the transverse
momentum by the \underline{virtuality} $Q$, to get to the standard
parameterization of parton splitting --- I know I am just waving my
hands at this stage, to understand the more fundamental role of the
virtuality we would have to look into deep inelastic scattering and
factorization. In terms of the virtuality, the splitting of one parton
into two is given by the splitting kernel integrated over the proper
range in the momentum fraction $x$
\begin{alignat}{5}
 d \mathcal{P}(x) &= \frac{\alpha_s}{2 \pi} \; 
                    \frac{d q^2}{q^2} \;
                    \int \; dx \; P(x)  
 \notag \\
 \mathcal{P}(Q_{\rm min},Q_{\rm max}) &= \frac{\alpha_s}{2 \pi} \; 
                              \int_{Q_{\rm min}}^{Q_{\rm max}} \;
                              \frac{d q^2}{q^2} \;
                              \int_{x_{\rm min}}^{x_{\rm max}} \;
                              dx \; P(x)  
\end{alignat}
The splitting kernel we symbolically write as $P(x)$, avoiding indices
and the sum over partons appearing in the DGLAP equation
eq.(\ref{eq:dglap}). The boundaries $x_{\rm min}$ and $x_{\rm max}$ we
can compute for example in terms of an over-all minimum value $Q_0$
and the actual values $q$, so we drop them for now. Strictly speaking,
the double integral over $x$ and $q^2$ can lead to two overlapping IR
divergences or logarithms, a soft logarithm arising from the $x$
integration (which we will not discuss further) and the collinear
logarithm arising from the virtuality integral. This is the logarithm
we are interested in when talking about the parton shower.

In the expression above we compute the probability that a parton will
split into another parton while moving from a virtuality $Q_{\rm max}$
down to $Q_{\rm min}$. This probability is given by QCD, as described
earlier. Using it, we can ask what the probability is that we will not
see a parton splitting from a parton starting at fixed $Q_{\rm max}$
to a variable scale $Q$, which is precisely the
\underline{Sudakov factor}
\begin{alignat}{5}
  \Delta(Q,Q_{\rm max}) &= 
  e^{- \mathcal{P} (Q,Q_{\rm max})} 
  \notag \\
  & =
  \exp \left[ - \frac{\alpha_s}{2 \pi}
                \int_Q^{Q_{\rm max}} \;
                \frac{d q^2}{q^2} \;
                \int_{x_{\rm min}}^{x_{\rm max}} \;
                dx \; P(x)
       \right] 
  \sim 
  e^{- \alpha_s \log^2 Q_{\rm max}/Q}
\end{alignat}
The last line omits all kinds of factors, but correctly identifies the
logarithms involved, namely $\alpha_s^n \log^{2n} Q_{\rm max}/Q$.

\subsection{Jet algorithm}

Before discussing methods to describe jets at the LHC we should
introduce one way to define jets in a detector, namely the
\underline{$k_T$ jet algorithm}. Imagine we observe a large number of
energy depositions in the calorimeter in the detector which we would
like to combine into jets.  We know that they come from a smaller
number of partons which originate in the hard QCD process and which
since have undergone a sizeable number of splittings. Can we try to
reconstruct partons?

The answer is yes, in the sense that we can combine a large number of
jets into smaller numbers, where unfortunately nothing tells us what
the final number of jets should be. This makes sense, because in QCD
we can produce an arbitrary number of hard jets in a hard matrix
element and another arbitrary number via collinear radiation. The main
difference between a hard jet and a jet from parton splitting is that
the latter will have a partner which originated from the same soft or
collinear splitting.\bigskip

The basic idea of the $k_T$ algorithm is to ask if a given jet has a
soft or collinear partner. For this we have to define a collinearity
measure, which will be something like the transverse momentum of one
jet with respect to another one $y_{ij} \sim k_{T, ij}$. If one of the
two jets is the beam direction, this measure simply becomes $y_{iB}
\sim k_{T,i}$. We define two jets as collinear, if $y_{ij} < y_{\rm
  cut}$ where $y_{\rm cut}$ we have to give to the algorithm. The jet
algorithm is simple:

\begin{itemize}
\item[(1)] for all final state jets find minimum 
         $y^{\rm min} = {\rm min}_{ij} (y_{ij},y_{iB})$
\item[(2a)] if $y^{\rm min} = y_{ij} < y_{\rm cut}$ merge jets $i$ and $j$, 
          go back to (1)
\item[(2b)] if $y^{\rm min} = y_{iB} < y_{\rm cut}$ remove jet $i$, 
          go back to (1)
\item[(2c)] if $y^{\rm min} > y_{\rm cut}$ keep all jets, done
\end{itemize}

The result of the algorithm will of course depend on the resolution
$y_{\rm cut}$. Alternatively, we can just give the algorithm the
minimum number of jets and stop there. The only question is what
`combine jets' means in terms of the 4-momentum of the new jet.  The
simplest thing would be to just combine the momentum vectors $k_i +
k_j \to k_i$, but we can still either combine the 3-momenta and give
the new jet a zero invariant mass (which assumes it indeed was one
parton) or we can add the 4-momenta and get a jet mass (which means
they can come from a $Z$, for example). But these are details for most
new-physics searches at the LHC. At this stage we run into a language
issue: what do we really call a jet? I am avoiding this issue by
saying that jet algorithms definitely start from calorimeter towers
and not jets and then move more and more towards jets, where likely
the last iterations could be described by combining jets into new
jets. \bigskip

From the QCD discussion above it is obvious why theorists prefer a
$k_T$ algorithm over for other algorithms which define the distance
between two jets in a more geometric manner: a jet algorithm combines
the complicated energy deposition in the hadronic calorimeter, and we
know that the showering probability or theoretically speaking the
collinear splitting probability is best described in terms of
virtuality or transverse momentum. A transverse-momentum distance
between jets is from a theory point of view best suited to combine the
right jets into the original parton from the hard
interaction. Moreover, this $k_T$ measure is intrinsically
infrared safe, which means the radiation of an additional soft parton
cannot affect the global structure of the reconstructed jets. For
other algorithms we have to ensure this property explicitly, and you
can find examples for this in QCD lectures by Mike Seymour.

One problem of the $k_T$ algorithm is that noise and the underlying
event can easiest be understood geometrically in the $4 \pi$
detector. Basically, the low-energy jet activity is constant all over
the detector, so the easiest thing to do is just subtract it from each
event. How much energy deposit we have to subtract from a
reconstructed jet depends on the actual area the jet covers in the
detector.  Therefore, it is a major step for the $k_T$ algorithm that
it can indeed compute an IR--safe geometric size of the jet. Even
more, if this size is considerably smaller than the usual geometric
measures, the $k_T$ algorithm should at the end of the day turn out to
be the best jet algorithm at the LHC.

\section{Jet merging} 

So how does a traditional Monte Carlo treat the radiation of jets
into the final state? It needs to reverse the summation of collinear
jets done by the DGLAP equation, because jet radiation is not strictly
collinear and does hit the detector. In other words, it computes
probabilities for radiating collinear jets from other jets and
simulates this radiation. Because it was the only thing we knew,
Monte Carlos used to do this in the collinear approximation.
However, from the brief introduction we know that at the LHC we should
generally not use the collinear approximation, which is one of the
reason why at the LHC we will use all-new Monte Carlos. Two ways how
they work we will discuss here. \bigskip

Apart from the collinear approximation for jet radiation, a second
problem with Monte Carlo simulation is that they `only do shapes'. In
other words, the normalization of the event sample will always be
perturbatively poorly defined. The simple reason is that collinear
jet radiation starts from a hard process and its production cross
section and from then on works with splitting probabilities, but never
touches the total cross section it started from.

Historically, people use higher-order cross sections to normalize the
total cross section in the Monte Carlo. This is what we call a
\underline{$K$ factor}: $K = \sigma^{\rm improved}/\sigma^{\rm MC} =
\sigma^{\rm improved}/\sigma^{\rm LO}$. It is crucial to remember that
higher-order cross sections integrate over unobserved additional jets
in the final state.  So when we normalize the Monte Carlo we assume
that we can first integrate over additional jets and obtain
$\sigma^{\rm improved}$ and then just normalize the Monte Carlo which
puts back these jets in the collinear approximation. Obviously, we
should try to do better than that, and there are two ways to improve
this traditional Monte Carlo approach.

\subsection{MC\@@NLO method}

When we compute the next-to-leading order correction to a cross
section, for example to Drell--Yan production, we consider all
contributions of the order $G_F \alpha_s$. There are three obvious
sets of Feynman diagrams we have to square and multiply, namely the
Born contribution $q \qb \to Z$, the virtual gluon exchange for
example between the incoming quarks, and the real gluon emission $q
\qb \to Zg$. Another set of diagrams we should not forget are the
crossed channels $q g \to Zq$ and $\qb g \to Z \qb$.  Only amplitudes
with the same external particles can be squared, so we get the
matrix-element-squared contributions
\begin{alignat}{5}
  |\mathcal{M}_B|^2 &\propto G_F \notag \\
  2 {\rm Re} \; \mathcal{M}_V^* \mathcal{M}_B &\propto G_F \alpha_s
   \quad \quad
  |\mathcal{M}_{Zg}|^2  &\propto G_F \alpha_s
   \quad \quad
  |\mathcal{M}_{Zq}|^2, |\mathcal{M}_{Z\qb}|^2 \propto G_F \alpha_s
\end{alignat}
Strictly speaking, we should have included the counter terms, which
are a modification of $|\mathcal{M}_B|^2$, shifted by counter terms of
the order $\alpha_s (1/\epsilon + C)$. These counter terms we add to
the interference of Born and virtual gluon diagrams to remove the UV
divergences. Luckily, this is not the part of the contributions we
want to discuss. IR poles can have two sources, soft and collinear
divergences. The first kind is cancelled between virtual gluon
exchange and real gluon emission. Again, we are not really interested
in them.

What we are interested in are the collinear divergences. They arise
from virtual gluon exchange as well as from gluon emission and from
gluon splitting in the crossed channels. The collinear limit is
described by the splitting kernels eq.(\ref{eq:splitting}), and the
divergences are absorbed in the re-definition of the parton densities
(like an IR pseudo-renormalization).\bigskip

To present the idea of MC\@@NLO Bryan Webber uses a nice toy model
which I am going to follow in a shortened version. It describes
simplified particle radiation off a hard process: the energy of the
system before radiation is $x_s$ and the energy of the outgoing
particle (call it photon or gluon) is $x$, so $x<x_s<1$. When we
compute \underline{next-to-leading order corrections} to a hard
process, the different contributions (now neglecting crossed channels)
are
\begin{equation}
            \frac{d \sigma}{dx} \Big|_B = B \, \delta(x) 
            \quad \quad
            \frac{d \sigma}{dx} \Big|_V = \alpha_s \left( \frac{B}{2\epsilon} + V
                                             \right) \delta(x) 
            \quad \quad
            \frac{d \sigma}{dx} \Big|_R = \alpha_s \frac{R(x)}{x} \, .\qquad
\end{equation}
The constant $B$ describes the Born process and the assumed
factorizing poles in the virtual contribution. The coupling constant
$\alpha_s$ should be extended by factors 2 and $\pi$, or color
factors.  We immediately see that the integral over $x$ in the real
emission rate is logarithmically divergent in the soft limit, similar
to the collinear divergences we now know and love.  From factorization
(\ie implying universality of the splitting kernels) we know that in
the collinear and soft limits the real emission part has to behave
like the Born matrix element $\lim_{x\to 0} R(x) = B$.\bigskip

The logarithmic IR divergence we extract in dimensional
regularization, as we already did for the virtual corrections. The
expectation value of any infrared safe observable over the entire
phase space is then given by
\begin{equation}
           \langle O \rangle
            = \mu_F^{2\epsilon}
              \int_0^1 \; dx \; \frac{O(x)}{x^{2\epsilon}}
                       \left[ \frac{d \sigma}{dx} \Big|_B
                             +\frac{d \sigma}{dx} \Big|_V
                             +\frac{d \sigma}{dx} \Big|_R
                       \right] \, .
\end{equation}
Dimensional regularization yields this additional factor
$1/x^{2\epsilon}$, which is precisely the factor whose mass unit we
cancel introducing the factorization scale $\mu_F^{2\epsilon}$. This
renormalization scale factor we will casually drop in the following.

When we compute a distribution of for example the energy of one of the
heavy particles in the process, we can extract a histogram from of the
integral for $\langle O \rangle$ and obtain a normalized distribution.
However, to compute such a histogram we have to numerically integrate
over $x$, and the individual parts of the integrand are not actually
integrable. To cure this problem, we can use the
\underline{subtraction method} to define integrable functions under
the $x$ integral. From the real emission contribution we subtract and
then add a smartly chosen term:
\begin{alignat}{5}
   \langle O \rangle_R
   =&  \int_0^1 \; dx \; \frac{O(x)}{x^{2\epsilon}} \;
                         \frac{d \sigma}{dx} \Big|_R 
   =   \int_0^1 \; dx \; \frac{O(x)}{x^{2\epsilon}} \;
                   \frac{\alpha_s R(x)}{x} \notag \\
   =&  \alpha_s \; B \; O(0) \int_0^1 \; dx \frac{1}{x^{1+2\epsilon}}
     + \int_0^1 dx \; \left( \frac{\alpha_s R(x)O(x)}{x^{1+2\epsilon}} 
                            -\frac{\alpha_s BO(0)}{x^{1+2\epsilon}}
                      \right) \notag  \\
   =&  \alpha_s \; B \; O(0) \int_0^1 \; dx \frac{1}{x^{1+2\epsilon}}
     + \alpha_s \int_0^1 dx \; \frac{R(x)O(x)-BO(0)}{x^{1+2\epsilon}} \notag \\
   =& -\alpha_s \frac{B \; O(0)}{2\epsilon}
     + \alpha_s \int_0^1 dx \; \frac{R(x)O(x)-BO(0)}{x}
\end{alignat}
In the second integral we take the limit $\epsilon \to 0$ because the
asymptotic behavior of $R(x \to 0)$ makes the numerator vanish and
hence regularizes this integral without any dimensional regularization
required. The first term precisely cancels the (soft) divergence from
the virtual correction. We end up with a perfectly finite $x$ integral
for all three contributions
\begin{alignat}{5}
\langle O \rangle
&= B \; O(0)
 + \alpha_s V \; O(0)
 + \alpha_s \int_0^1 \; dx \; \frac{R(x)\; O(x) -B \; O(0)}{x}
 \notag \\
&= \int_0^1 \; dx \; 
 \left[ 
        O(0) \; \left( B + \alpha_s V - \alpha_s \frac{B}{x} \right) 
      + O(x) \; \alpha_s \; \frac{R(x)}{x} 
 \right]
\end{alignat}
This procedure is one of the standard methods to compute
next-to-leading order corrections involving one-loop virtual
contributions and the emission of one additional parton. This formula
is a little tricky: usually, the Born-type kinematics would come with
an explicit factor $\delta(x)$, which in this special case we can omit
because of the integration boundaries. We can re-write the same
formula in terms of a derivative
\begin{equation}
 \frac{d \sigma}{d O} 
= \int_0^1 \; dx \; 
 \left[ I(O)_{\rm LO} \;
        \left( B + \alpha_s V - \alpha_s \frac{B}{x} 
        \right) 
      + I(O)_{\rm NLO} \;
        \alpha_s \; \frac{R(x)}{x} 
 \right]
\label{eq:naive}
\end{equation}
The transfer function $I(O)$ is defined in a way that formally does
precisely what we did before: at leading order we evaluate it using
the Born kinematics $x=0$ while allowing for a general $x=0 \cdots 1$
for the real emission kinematics.\bigskip

In this calculation we have integrated over the entire phase space of
the additional parton. For a hard additional parton or jet everything
looks well defined and finite. On the other hand, we cancel an IR
divergence in the virtual corrections proportional to a Born-type
momentum configuration $\delta(x)$ with another IR divergence which
appears after integrating over small but finite values of $x \to
0$. In a histogram in $x$, where we encounter the real-emission
divergence at small $x$, this divergence is cancelled by a negative
delta distribution right at $x=0$.  Obviously, this will not give us a
well-behaved distribution.  What we would rather want is a way to
smear out this pole such that it coincides with the in that range
justified collinear approximation and cancels the real emission over
the entire low-$x$ range. At the same time it has to leave the hard
emission intact and when integrated give the same result as the
next-to-leading oder rate. Such a modification will use the emission
probability or Sudakov factors. We can define an emission probability
of a particle with an energy fraction $z$ as $d \mathcal{P} = \alpha_s
E(z)/z \, dz$. Note that we have avoided the complicated proper
two--dimensional description in favor of this simpler picture just in
terms of particle energy fractions.\bigskip

Let us consider a perfectly fine observable, the radiated photon
spectrum as a function of the (external) energy scale $z$. We know
what this spectrum has to look like for the two kinematic
configurations
\begin{equation}
 \frac{d \sigma}{d z} \Big|_{\rm LO} = \alpha_s \; \frac{B E(z)}{z}
 \qquad \qquad 
 \frac{d \sigma}{d z} \Big|_{\rm NLO} = \alpha_s \; \frac{R(z)}{z} 
\end{equation}
The first term corresponds to parton shower radiation from the Born
diagram (at order $\alpha_s$), while the second term is the real
emission defined above. The transfer functions we would have to
include in eq.(\ref{eq:naive}) to arrive at this equation for the
observable are
\begin{alignat}{5}
 I(z,1) \Big|_{\rm LO} &= \alpha_s \; \frac{E(z)}{z}
 \notag \\
 I(z,x_M) \Big|_{\rm NLO} &= \delta(z-x) 
                      + \alpha_s \; \frac{E(z)}{z} \; \Theta(x_M(x)-z)
\end{alignat}
The additional second term in the real-radiation transfer function
arises from a parton shower acting on the real emission process. It
explicitly requires that enough energy has to be available to radiate
a photon with an energy $z$, where $x_M$ is the energy available at
the respective stage of showering, \ie $z < x_M$.

These transfer functions we can include in eq.(\ref{eq:naive}), which
becomes
\begin{alignat}{5}
 \frac{d \sigma}{d z} 
&= \int_0^1 dx \; 
 \left[ I(z,1) \;
        \left( B + \alpha_s V - \alpha_s \frac{B}{x} 
        \right) 
      + I(z,x_M) \;
        \alpha_s \; \frac{R(x)}{x} 
 \right] \notag \\
&= \int_0^1 dx \; 
 \left[ \alpha_s \frac{E(z)}{z}
        \left( B + \alpha_s V - \alpha_s \frac{B}{x}
        \right)
      + \left( \delta(x-z) + \mathcal{O} (\alpha_s)
        \right)
        \; \alpha_s \frac{R(x)}{x} 
   \right] \notag \\
&= \int_0^1 dx \; 
 \left[ \alpha_s \; \frac{B E(z)}{z}
      + \alpha_s \; \frac{R(z)}{z} 
   \right] + \mathcal{O} (\alpha_s^2) \notag \\
&= \alpha_s \; \frac{B E(z) + R(z)}{z}
 + \mathcal{O} (\alpha_s^2)
\end{alignat}
All Born--type contributions proportional to $\delta(z)$ have vanished
by definition. This means we should be able to integrate the $z$
distribution to the total cross section $\sigma_{\rm tot}$ with a
$z_{\rm min}$ cutoff for consistency. However, the distribution we
obtained above has an additional term which spoils this agreement, so
we are still missing something.\bigskip

On the other hand, we also knew we would fall short, because what we
described in words about a subtraction term for finite $x$ cancelling
the real emission we have not yet included. This means, first we have
to add a subtraction term to the real emission which cancels the
fixed-order contributions for small $x$ values. Because of
factorization we know how to write such a subtraction term using the
splitting function, called $E$ in this example:
\begin{equation}
\frac{R(x)}{x} \quad \longrightarrow \quad
\frac{R(x) - B E(x)}{x}
\end{equation}
To avoid double counting we have to add this parton shower to the
Born-type contribution, now in the collinear limit, which leads us to
a modified version of eq.(\ref{eq:naive})
\begin{alignat}{5}
  \frac{d \sigma}{d O} = \int_0^1 \; dx \; 
    \Bigg[\quad &I(O,1) \; \left( B 
                        + \alpha_s V 
                        - \frac{\alpha_s B}{x} 
                        + \frac{\alpha_s B E(x)}{x} 
                   \right) 
  \notag \\
       +&I(O,x_M) \; \alpha_s \frac{R(x)-B E(x)}{x}
  \Bigg] 
\end{alignat}
When we again compute the $z$ spectrum to order $\alpha_s$ there will
be an additional contribution from the Born-type kinematics
\begin{alignat}{5}
 \frac{d \sigma}{d z} 
&= \int_0^1 \; dx \; \alpha_s \; \frac{B E(z) + R(z)}{z} 
    + \mathcal{O} (\alpha_s^2) \notag \\
&\longrightarrow 
   \int_0^1 \; dx \; \left[ \alpha_s \; \frac{B E(z) + R(z)}{z}
                          - \alpha_s \; \delta(x-z) \; \frac{B E(x)}{x} 
                     \right] 
    + \mathcal{O} (\alpha_s^2) \notag \\
&= \int_0^1 \; dx \; \alpha_s \; \frac{B E(z) + R(z) - B E(z)}{z} 
    + \mathcal{O} (\alpha_s^2) \notag \\
&= \alpha_s \; \frac{R(z)}{z}  + \mathcal{O} (\alpha_s^2)
\end{alignat}
which gives us the distribution we expected, without any double
counting.\bigskip

In other words, this scheme implemented in the MC\@@NLO Monte Carlo
describes the hard emission just like a next-to-leading order
calculation, including the next-to-leading order normalization.  On
top of that, it simulates additional collinear particle emissions
using the Sudakov factor. This is precisely what the parton shower
does. Most importantly, it avoids double counting between the first
hard emission and the collinear jets, which means it describes the
entire $p_T$ range of jet emission for the \underline{first and
hardest} radiated jet consistently. Additional jets, which do not
appear in the next-to-leading order calculation are simply added by
the parton shower, \ie in the collinear approximation. What looked to
easy in our toy example is of course much harder in the mean QCD
reality, but the general idea is the same: to combine a fixed-order
NLO calculation with a parton shower one can think of the parton
shower as a contribution which cancels a properly defined subtraction
term which we can include as part of the real emission contribution.

\subsection{CKKW method}

The one weakness of the MC\@@NLO method is that it only describes one
hard jet properly and relies on a parton shower and its collinear
approximation to simulate the remaining jets. Following the general
rule that there is no such thing as a free lunch we can improve on the
number of correctly described jets, which unfortunately will cost us
the next-to-leading order normalization.\bigskip

For simplicity, we will limit our discussion to final state
radiation, for example in the inverse Drell--Yan process $e^+ e^- \to
q \qb$. We know already that this final state is likely to evolve into
more than two jets. First, we can radiate a gluon off one of the quark
legs, which gives us a $q \qb g$ final state, provided our $k_T$
algorithm finds $y_{ij} > y_{\rm cut}$. Additional splittings can also
give us any number of jets, and it is not clear how we can combine
these different channels.

Each of these processes can be described either using matrix elements
or using a parton shower, where `describe' means for example compute
the relative probability of different phase space configurations. The
parton shower will do well for jets which are fairly collinear,
$y_{ij} < y_{\rm ini}$. In contrast, if for our closest jets we find
$y_{ij} > y_{\rm ini}$, we know that collinear logarithms did not play
a major role, so we can and should use the hard matrix element. How do
we combine these two approaches?

The CKKW scheme tackles this multi-jet problem. It first allows us to
combine final states with a \underline{different number of jets}, and
then ensures that we can add a parton shower without any double
counting. The only thing I will never understand is that they labelled
the transition scale as `ini'.\bigskip

Using Sudakov factors we can first construct the
\underline{probabilities} of generating $n$--jet events from a hard
two--jet production process. These probabilities make no assumptions
on how we compute the actual kinematics of the jet radiation, \ie if
we model collinear jets with a parton shower or hard jets with a
matrix element.  This way we will also get a rough idea how Sudakov
factors work in practice. For the two--jet and three--jet final
states, we will see that we only have to consider the splitting
probabilities for the different partons
\begin{alignat}{5}
  \Gamma_q(Q_{\rm out}, Q_{\rm in}) 
         &\equiv \Gamma_{q \leftarrow q}(Q_{\rm out}, Q_{\rm in})
           = \frac{2 C_F}{\pi} \; \frac{\alpha_s(Q_{\rm out})}{Q_{\rm out}} \;
             \left( \log \frac{Q_{\rm in}}{Q_{\rm out}} 
                  - \frac{3}{4} 
             \right) \notag \\
  \Gamma_g(Q_{\rm out}, Q_{\rm in}) 
         &\equiv \Gamma_{g \leftarrow q}(Q_{\rm out}, Q_{\rm in})
           = \frac{2 C_A}{\pi} \; \frac{\alpha_s(Q_{\rm out})}{Q_{\rm out}} \;
             \left( \log \frac{Q_{\rm in}}{Q_{\rm out}} 
                  - \frac{11}{12} 
             \right) 
\label{eq:split_const}
\end{alignat}
The virtualities $Q_{\rm in, out}$ correspond to the incoming (mother)
and outgoing (daughter) parton. Unfortunately, this formula is
somewhat understandable from the argument before and from $P_{q
  \leftarrow q}$, but not quite. That has to do with the fact that
these splittings are not only collinearly divergent, but also softly
divergent, as we can see in the limits $x \to 0$ and $x \to 1$ in
eq.(\ref{eq:splitting}). These divergences we have to subtract first,
so the formulas for the splitting probabilities $\Gamma_{q,g}$ look
unfamiliar. In addition, we find finite terms arising from
next-to-leading logarithms which spoil the limit $Q_{\rm out} \to
Q_{\rm in}$, where the probability of no splitting should go to
unity. But at least we can see the leading (collinear) logarithm $\log
Q_{\rm in}/Q_{\rm out}$. Technically, we can deal with the finite
terms in the Sudakov factors by requiring them to be positive
semi-definite, \ie by replacing $\Gamma(Q_{\rm out}, Q_{\rm in}) < 0$
by zero.

Given the splitting probabilities we can write down the
\underline{Sudakov factor}, which is the probability of not radiating
any hard and collinear gluon between the two virtualities:
\begin{equation}
  \Delta_{q,g}(Q_{\rm out}, Q_{\rm in})
  = \exp \left[ - \int_{Q_{\rm out}}^{Q_{\rm in}} \; dq \;
                  \Gamma_{q,g}(q, Q_{\rm in})
         \right]
\end{equation}
This integral boundaries are $Q_{\rm out} < Q_{\rm in}$. This
description we can generalize for all splittings $P_{i \leftarrow j}$
we wrote down before.\bigskip

First, we can compute the probability that we see exactly
\underline{two partons}, which means that none of the two quarks radiate
a resolved gluon between the virtualities $Q_2$ and $Q_1$, where we
assume that $Q_1 < Q_2$ gives the scale for this resolution. It is
simply $\left[ \Delta_q(Q_1, Q_2) \right]^2$, once for each quark, so
that was easy.\bigskip

Next, what is the probability that the two--jet final state evolves
exactly into \underline{three partons}? We know that it contains a
factor $\Delta_q(Q_1, Q_2)$ for one untouched quark. If we label the
point of splitting in the matrix element $Q_q$ for the quark, there
has to be a probability for the second quark to get from $Q_2$ to
$Q_q$ untouched, but we leave this to later. After splitting with the
probability $\Gamma_q(Q_2,Q_q)$, this quark has to survive to $Q_1$,
so we have a factor $\Delta_q(Q_1, Q_q)$.  Let's call the virtuality
of the radiated gluon after splitting $Q_g$, then we find the gluon's
survival probability $\Delta_g(Q_1,Q_g)$. So what we have until now is
\begin{equation}
 \Delta_q(Q_1, Q_2) \;
 \Gamma_q(Q_2,Q_q) \;
 \Delta_q(Q_1, Q_q) \;
 \Delta_g(Q_1,Q_g) \cdots
\end{equation}
That's all there is, with the exception of the intermediate quark.
Naively, we would guess its survival probability between $Q_2$ and
$Q_q$ to be $\Delta_q(Q_q,Q_2)$, but that is not correct. That would
imply no splittings resolved at $Q_q$, but what we really mean is no
splitting resolved later at $Q_1 < Q_q$.  Instead, we compute the
probability of no splitting between $Q_2$ and $Q_q$ from
$\Delta_q(Q_1,Q_2)$ under the additional condition that splittings
from $Q_q$ down to $Q_1$ are now allowed. If no splitting occurs
between $Q_1$ and $Q_q$ this simply gives us $\Delta_q(Q_1,Q_2)$ for
the Sudakov factor between $Q_2$ and $Q_q$. If one splitting happens
after $Q_q$ this is fine, but we need to add this combination to the
Sudakov between $Q_2$ and $Q_q$. Allowing an arbitrary number of
possible splittings between $Q_q$ and $Q_1$ gives us
\begin{alignat}{5}
 &\Delta_q(Q_1,Q_2) \left[ 1 
                         + \int_{Q_q}^{Q_1} dq \; \Gamma_q(q,Q_1)
                         + \cdots 
                   \right] =
  \notag \\
 & \qquad \qquad \quad = \Delta_q(Q_1,Q_2) \;
   \exp \left[  \int_{Q_q}^{Q_1} dq \; \Gamma_q(q,Q_1) \right]
 = \frac{\Delta_q(Q_1,Q_2)}{\Delta_q(Q_1,Q_q)} \; .
\label{eq:veto}
\end{alignat}
So once again: the probability of nothing happening between $Q_2$ and
$Q_q$ we compute from the probability of nothing happening between
$Q_2$ and $Q_1$ times possible splittings between $Q_q$ and
$Q_1$. \bigskip

Collecting all these factors gives the combined probability that we
find exactly three partons at a virtuality $Q_1$
\begin{alignat}{5}
  \Delta_q(Q_1,Q_2) \;
  \Gamma_q(Q_2,Q_q) \; 
  \Delta_q(Q_1,Q_q) \; 
  \Delta_g(Q_1,Q_g) \; 
  \frac{\Delta_q(Q_1,Q_2)}{\Delta_q(Q_1,Q_q)} \notag \\
 = \Gamma_q(Q_2,Q_q)  \; [\Delta_q(Q_1,Q_2)]^2 \; \Delta_g(Q_1,Q_g)
\end{alignat}
This result is pretty much what we would expected: both quarks go
through untouched, just like in the two--parton case. But in addition
we need exactly one splitting producing a gluon, and this gluon cannot
split further. This example illustrates how it is fairly easy to
compute these probabilities using Sudakov factors: adding a gluon
corresponds to adding a splitting probability times the survival
probability for this gluon, everything else magically drops out. At
the end, we only integrate over the splitting point $Q_q$.\bigskip

The first part of the CKKW scheme we illustrate is how to combine
different $n$--parton channels in one framework.  Knowing some of the
basics we can write down the (simplified) \underline{CKKW algorithm}
for final state radiation. As a starting point, we compute all
leading-order cross sections for $n$-jet production with a lower
cutoff at $y_{\rm ini}$.  This cutoff ensures that all jets are hard
and that all $\sigma_{n,i}$ are finite. The second index $i$ describes
different non-interfering parton configurations, like $q\qb gg$ and
$q\qb q\qb$ for $n=4$.  The purpose of the algorithm is to assign a
weight (probability, matrix element squared,...) to a given
phase space point, statistically picking the correct process and
combining them properly.

\begin{itemize}
\item[(1)] for each jet final state $(n,i)$ compute the relative
  probability $P_{n,i} = \sigma_{n,i}/\sum \sigma_{k,j}$; select a
  final state with this probability $P_{n,i}$
\item[(2)] distribute the jet momenta to match the external particles
  in the matrix element and compute $\MX$
\item[(3)] use the $k_T$ algorithm to compute the virtualities $Q_j$
  for each splitting in this matrix element
\item[(4)] for each internal line going from $Q_j$ to $Q_k$ compute
  the Sudakov factor $\Delta(Q_1,Q_j)/\Delta(Q_1,Q_k)$, where $Q_1$ is
  the final resolution of the evolution. For any final state line
  starting at $Q_j$ apply $\Delta(Q_1,Q_j)$. All these factors
  combined give the combined survival probability described above.
\end{itemize}

The matrix element weight times the survival probability can be used
to compute distributions from weighted events or to decide if to keep
or discard an event when producing unweighted events. The line of
Sudakov factors ensures that the relative weight of the different
$n$--jet rates is identical to the probabilities we just computed.
Their kinematics, however, are hard--jet configuration without any
collinear assumption. There is one remaining subtlety in this
procedure which I am skipping.  This is the re-weighting of
$\alpha_s$, because the hard matrix element will be typically computed
with a fixed hard renormalization scale, while the parton shower only
works with a scale fixed by the virtuality of the respective
splitting. But those are details, and there will be many more details
in which different implementations of the CKKW scheme differ.\bigskip

The second question is what we have to do to match the hard matrix
element with the parton shower at a critical resolution point $y_{\rm
  ini} = Q_1^2/Q_2^2$. From $Q_1$ to $Q_0$ we will use the parton
shower, but above this the matrix elements will be the better
description. For both regimes we already know how to combine different
$n$--jet processes. On the other hand, we need to make sure that this
last step does not lead to any double counting. From the discussion
above, we know that Sudakovs which describe the evolution between
scales but use a lower virtuality as the resolution point are going to
be the problem. On the other hand, we also know how to describe this
behavior using the additional splitting factors we used for the $Q_2
\cdots Q_q$ range. Carefully distinguishing the virtuality scale of
the actual splitting and the scale of jet resolution is the key, which
we have to combine with the fact that in the CKKW method starts each
parton shower at the point where the parton first appears. It turns
out that we can use this argument to keep the resolution ranges $y >
y_{\rm ini}$ and $y < y_{\rm ini}$ separate, without any double
counting. There is a simple way to check this, namely the question if
the \underline{$y_{\rm ini}$ dependence} drops out of the final
combined probabilities. And the answer for final state radiation is
yes, as proven in the original paper, including a hypothetical
next-to-leading logarithm parton shower.\bigskip

One widely used variant of CKKW is Michelangelo Mangano's
\underline{MLM scheme}, for example implemented in Alpgen or
Madevent. Its main difference to the classical CKKW is that it avoids
computing the corresponding survival properties using Sudakov form
factors. Instead, it vetoes events which CKKW would have cut using the
Sudakov rescaling.  This way it avoids problems with splitting
probabilities beyond the leading logarithms, for example the finite
terms appearing in eq.(\ref{eq:split_const}) which can otherwise lead
to a mismatch between the actual shower evolution and the analytic
expressions of the Sudakov factors.  Its veto approach allows the MLM
scheme to combine a set of $n$--parton events after they have been
generated using hard matrix elements. Its parton shower is then not
needed to compute a Sudakov reweighting. On the other hand, to combine
a given sample of events the parton shower has to start from an
external scale, which should be chosen as the hard(est) scale of the
process.

Once the parton shower has defined the complete event, we need to
decide if this event needs to be removed to avoid double counting due
to an overlap of simulated collinear and hard radiation. After
applying a jet algorithm (which in the case of Alpgen is a cone
algorithm and in case of Madevent is a $k_T$ algorithm) we can simply
compare the hard event with the showered event by identifying each
reconstructed showered jet with the partons we started from. If all
jet--parton combinations match and there are not additional resolved
jets apart from the highest-multiplicity sample we know that the
showering has not altered the hard-jet structure of the event,
otherwise the event has to go.

Unfortunately, the vetoing approach does not completely save the MLM
scheme the backwards evolution of a generated event, since we still
need to know the energy or virtuality scales at which partons split to
fix the scale of the strong coupling. If we know the Feynman diagrams
which lead to each event, we can check that a certain splitting is
actually possible in its color structure.

In my non-expert user's mind, all merging schemes are conceptually
similar enough that we should expect them to reproduce each others'
results, and they largely do. But the devil is in the details, and we
have to watch out for example for threshold kinks in jet distributions
which should not be there. \bigskip

\begin{figure}[t]
\begin{center}
\includegraphics[width=6.5cm,angle=0]{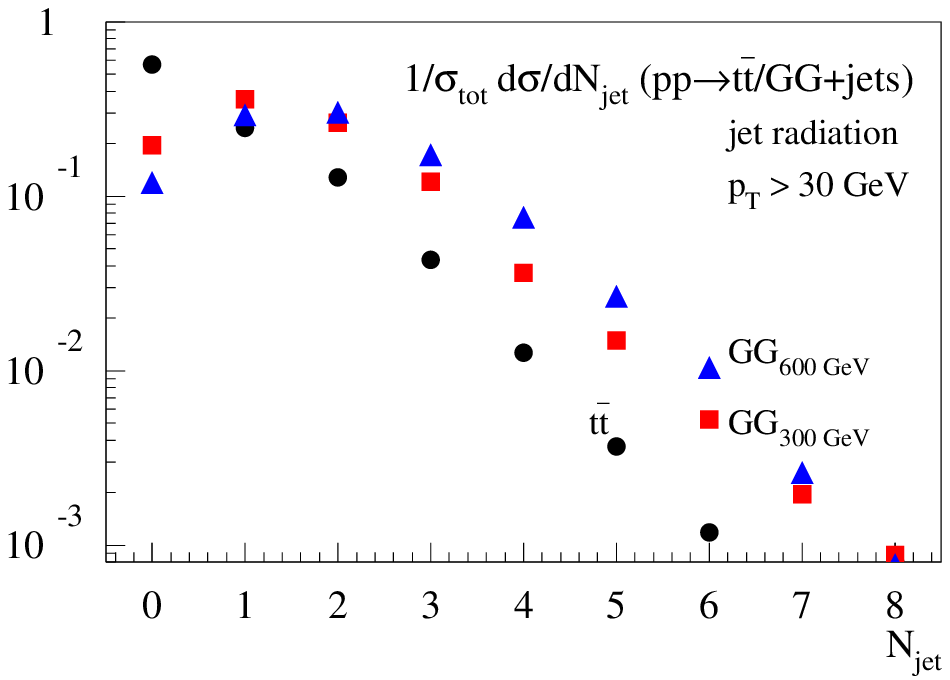}
 \hspace*{2cm}
\includegraphics[width=6.5cm,angle=0]{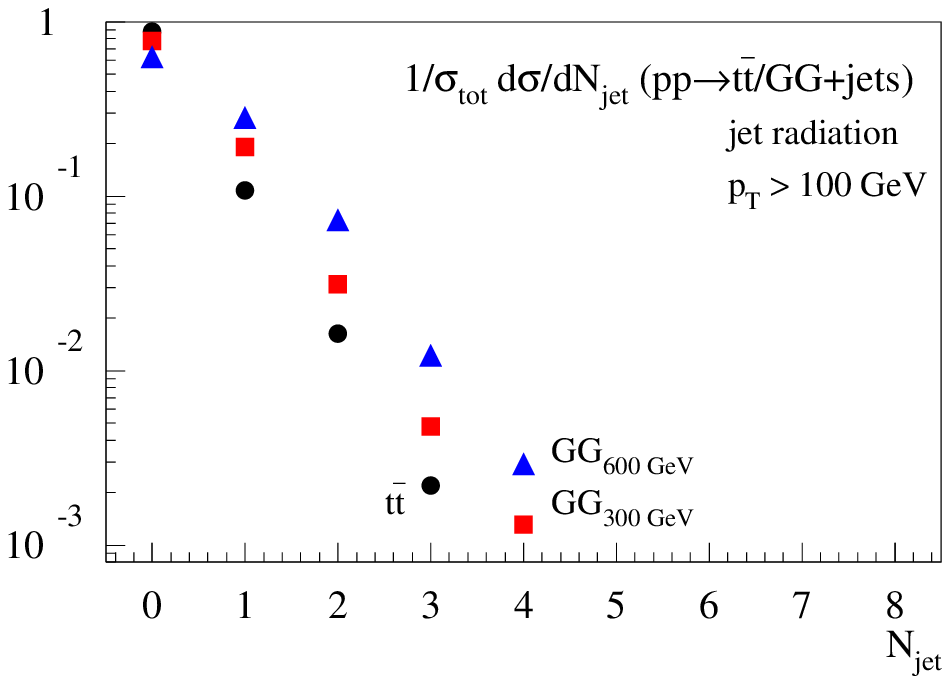}
\end{center}
\vspace*{-0.5cm}
\caption{Number of additional jets with a transverse momentum of at
  least 30 or 100~GeV radiated from top pair production and the
  production of heavy states at the LHC. As an example for such heavy
  states we use a pair of scalar gluons with a mass of 300 or 600~GeV,
  pair-produced in gluon fusion. The figures are from a forthcoming
  paper with Tim Tait (arXiv:0810:3919), produced with MadEvent using
  it's modified MLM algorithm --- thanks to Johan Alwall.}
\label{fig:madgraph}
\end{figure}

\begin{table}[b]
\begin{tabular}{l||l|l}
   & MC\@@NLO (Herwig)
   & CKKW (Sherpa) \\ \hline 
hard jets
   & first jet correct
   & all jets correct \\
collinear jets
   & all jets correct, tuned
   & all jets correct, tuned \\
normalization
   & correct to NLO
   & correct to LO plus real emission \\
variants
   & Powheg,...
   & MLM--Alpgen, MadEvent,... \\
\end{tabular}
\caption{Comparison of the MC\@@NLO and CKKW schemes combining
  collinear and hard jets.}
\label{tab:merging}
\end{table}

To summarize, we can use the CKKW or MLM schemes to combine $n$-jet
events with variable $n$ and at the same time combine matrix element
and parton shower descriptions of the jet kinematics. In other words,
we can for example simulate $Z+n$~jets production at the LHC, where
all we have to do is cut off the number of jets at some point where we
cannot compute the matrix element anymore. This combination will
describe all jets correctly over the entire collinear and hard phase
space. In Fig.\ref{fig:madgraph} we show the number of jets produced
in association with a pair of top quarks and a pair of heavy new
states at the LHC. The details of these heavy scalar gluons are
secondary for the basic features of these distributions, the only
parameter which matters is their mass, \ie the hard scale of the
process which sets the factorization scale and defines the upper limit
of collinearly enhanced initial-state radiation. We see that heavy
states tend to come with several jets radiated with transverse momenta
up to 30~GeV, where most of these jets vanish once we require
transverse momenta of at least 100~GeV. Looking at this figure you can
immediately see that a suggested analysis which for example asks for a
reconstruction of two $W$ decay jets better give you a very good
argument why it should not we swamped by combinatorics.

Looking at the individual columns in Fig.\ref{fig:madgraph} there is
one thing we have to keep in mind: each of the merged matrix elements
combined into this sample is computed at leading order, the emission
of real particles is included, while virtual corrections are not
(completely) there. In other words, in contrast to MC\@@NLO this
procedure gives us all jet distributions but leaves the normalization
free, just like an old-fashioned Monte Carlo. The main features and
shortcomings of the two merging schemes are summarized in
Tab.\ref{tab:merging}. A careful study of the associated theory errors
for example for $Z$+jets production and the associated rates and
shapes I have not yet come across, but watch out for it.\bigskip

As mentioned before --- there is no such thing as a free lunch, and it
is up to the competent user to pick the scheme which suits their
problem best. If there is a well-defined hard scale in the process,
the old-fashioned Monte Carlo with a tuned parton shower will be
fine, and it is by far the fastest method.  Sometimes we are only
interested in one hard jet, so we can use MC\@@NLO and benefit from
the correct normalization. And in other cases we really need a large
number of jets correctly described, which means CKKW and some external
normalization. This decision is not based on chemistry, philosophy or
sports, it is based on QCD. What we LHC phenomenologists have to do is
to get it right and know why we got it right.

On the other hand I am not getting tired of emphasizing that the
conceptual progress in QCD describing jet radiation for all
transverse-momentum scales is absolutely crucial for LHC analyses. If
I were a string theorist I would definitely call this achievement a
revolution or even two, like 1917 but with the trombones and cannons
of Tchaikovsky's 1812. In contrast to a lot of progress in theoretical
physics jet merging solves a very serious problem which would have
limited our ability to understand LHC data, no matter what kind of
Higgs or new physics we are looking for. And I am not sure if I got
the message across --- the QCD aspects behind it are not trivial at
all. If you feel like looking at a tough problem, try to prove that
CKKW and MLM work for initial-state and final-state
radiation... \bigskip

\begin{figure}[t]
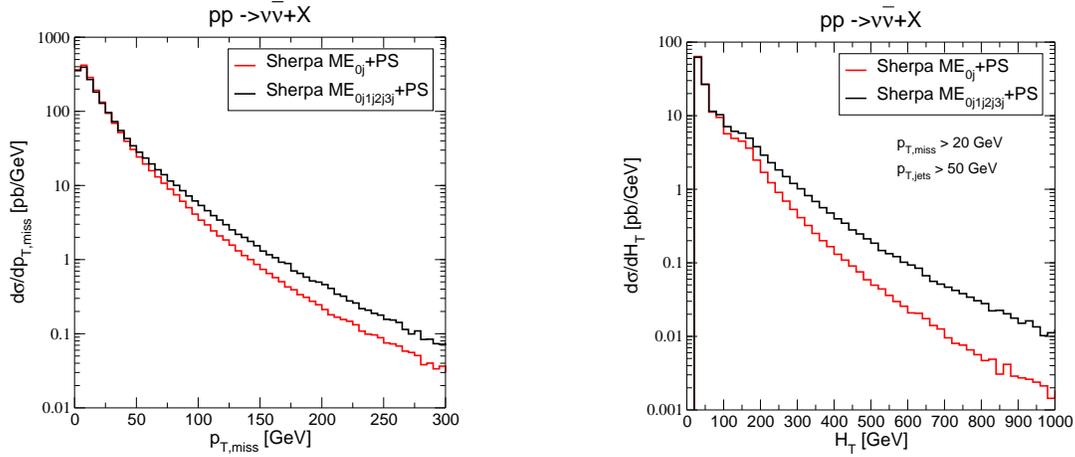

\begin{center}
\includegraphics[width=6cm,angle=0]{Sherpa_PTmiss.eps}
 \hspace*{2cm}
\includegraphics[width=6cm,angle=0]{Sherpa_HT_50GeV.eps}
\end{center}
\vspace*{-0.5cm}
\caption{Transverse momentum and $H_T$ distributions for $Z$+jets
  production at the LHC. The two curves correspond to the Sherpa
  parton shower starting from Drell--Yan production and the fully
  merged sample including up to three hard jets. These distributions
  describe typical backgrounds for searches for jets plus missing
  energy, which could originate in supersymmetric squark and gluino
  production. Thank you to Steffen Schumann and Sherpa for providing
  these Figures.}
\label{fig:ckkw}
\end{figure}

Before we move on, let me illustrate why in Higgs or exotics searches
at the LHC we really care about this kind of progress in QCD.  One way
to look for heavy particles decaying into jets, leptons and missing
energy is the variable
\begin{alignat}{5}
 H_T &= \sla{E}_T + \sum_j E_{T,j} + \sum_\ell E_{T,\ell} 
 \notag \\
     &= \sla{p}_T + \sum_j p_{T,j} + \sum_\ell p_{T,\ell}
 \qquad \text{(for massless quarks, leptons)}
\end{alignat}
which for gluon-induced QCD processes should be as small as possible,
while the signal's scale will be determined by the new particle
masses. For the background process $Z$+jets, this distribution as well
as the missing energy distribution using CKKW as well as a parton
shower (both from Sherpa) are shown in Fig.~\ref{fig:ckkw}. The two
curves beautifully show that the naive parton shower is not a good
description of QCD background processes to the production of heavy
particles. We can probably use a chemistry approach and tune the
parton shower to correctly describe the data even in this parameter
region, but we would most likely violate basic concepts like
factorization. How much you care about this violation is up to you,
because we know that there is a steep gradient in theory standards
from first-principle calculations of hard scattering all the way to
hadronization string models...\bigskip

\section{Simulating LHC events}

In the third main section I will try to cover a few topics of interest
to LHC physicists, but which are not really theory problems. Because
they are crucial for our simulations of LHC signatures and can turn
into sources of great embarrassment when we get them wrong in public.

\subsection{Missing energy} 

\begin{figure}[t]
\begin{center}
\includegraphics[width=7cm,angle=0]{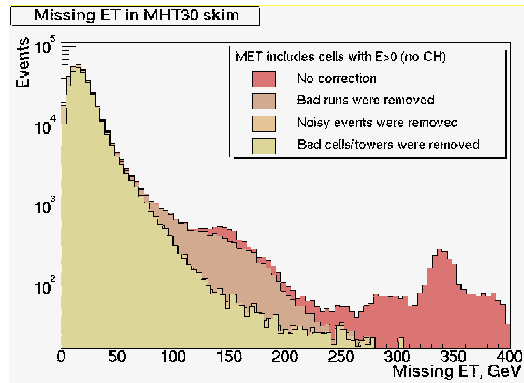}
\end{center}
\vspace*{-0.5cm}
\caption{Missing energy distribution from the early running phase of
  the DZero experiment at the Tevatron. This figure I got from Beate
  Heinemann's lectures web site.}
\label{fig:met}
\end{figure}

Some of the most interesting signatures at the LHC involve
dark matter particles. Typically, we would produce strongly
interacting new particles which then decay to the weakly interacting
dark matter agent. On the way, the originally produced particles have
to radiate quarks or gluons, to get rid of their color charge. If they
also radiate leptons, those can be very useful to trigger on the
events and reduce QCD backgrounds.

At the end of the last section we talked about the proper simulation
of $W$+jets and $Z$+jets backgrounds to such signals. It turns out
that jet merging predicts considerably larger missing transverse
momentum from QCD sources, so theoretically we are on fairly safe
ground. However, this is not the whole story of missing transverse
momentum. I should say that I skipped most of this section, because
Peter Wittich knows much more about it and covered it really
nicely. But it might nevertheless be useful to include it in this
writeup.\bigskip

Fig.~\ref{fig:met} is a historic missing transverse energy
distribution from DZero. It nicely illustrates that by just measuring
missing energy, Tevatron would have discovered supersymmetry with two
beautiful peaks in the missing-momentum distribution around 150~GeV
and around 350~GeV. However, this distribution has nothing to do with
physics, it is purely a detector effect.

The problem of missing energy can be illustrated with a simple number:
to identify and measure a lepton we need around 500 out of 200000
calorimeter cells in an experiment like Atlas, while for missing
energy we need all of them. Therefore, we need to understand our
detectors really well to even cut on a variable like missing
transverse momentum, and for this level of understanding we need time
and luminosity. Unless something goes wrong with the machine, I would
not expect us to find anything reasonable in early-running LHC data
including a missing energy cut --- really, we should not use the
phrases `missing energy' and `early running' in the same sentences or
papers.\bigskip

There are three sources of missing energy which our experimental
colleagues have to understand before we get to look at such
distributions:

First, we have to subtract bad runs. This means that for a few hours
parts of the detector might not have worked properly. We can identify
such bad runs by looking at Standard Model physics, like gauge bosons,
and remove them from the data sample.

Next, there is usually coherent noise in the calorimeter. Of 200000
cells we know that some of them will individually fail or produce
noise. However, some sources of noise, like leaking voltage or other
electronic noise can be correlated geometrically, \ie coherent. Such
noise will lead to beautiful missing momentum signals.  In the same
spirit, there might also be particles crossing our detector, but not
coming from the interaction point. Such particles can be cosmic rays
or errand beam radiation, and they will lead to unbalanced energy
deposition in the calorimeter. The way to get rid of such noise is
again looking for Standard Model candles and remove sets of events
where such problems occur.

The third class of fake missing energy is failing calorimeter cells,
like continuously hot cells or dead cells, which can be removed after
we know the detector really well.\bigskip

Once we understand all the source of fake missing momentum we can
focus on real missing momentum. This missing transverse momentum is
trivially computed from the momentum measurement of all tracks seen in
the detector. This means that any uncertainty on these measurements,
like the jet or lepton energy scale will smear the missing
momentum. Moreover, we know that there is for example dead matter in
the detector, so we have to compensate for this. This compensation is
obviously a global correction to individual events, which means it
will generally smear the missing energy distribution. So when we
compute a realistic missing transverse momentum distribution at the
LHC we have to smear all jet and lepton momenta, and in addition apply
a Gaussian \underline{smearing} of the order
\begin{equation}
\frac{\Delta \sla{E}_T}{\text{GeV}}
\sim 
\frac{1}{2} \; \sqrt{ \frac{\sum E_T}{\text{GeV}}} \;
\gtrsim 20
\end{equation}
While this sounds like a trivial piece of information I cannot count
the number of papers I get to referee where people forgot this
smearing and discovered great channels to look for Higgs bosons or new
physics at the LHC which completely fall apart when experimentalists
take a careful look. Here comes another great piece of phenomenology
wisdom: phenomenological studies are right or wrong based on the
outcome if they can be reproduced by real experimentalists and real
detectors --- at least once we make sure our experimentalist friends
did not screw it up again....

\subsection{Phase space integration} 

At the very beginning of this lecture we discussed how to compute the
total cross section for interesting processes. What we skipped is how
to numerically compute such cross sections. Obviously, since the
parton densities are not known in a closed analytical form, we will
have to rely on numerical integration tools. Looking at a simple $2
\to 2$ process we can write the total cross section as 
\begin{equation}
\sigma_{\rm tot} = \int d \phi 
                   \int d \cos \theta
                   \int d x_1
                   \int d x_2 \;
                   F_{\rm PS} 
                   \left| \mathcal{M} \right|^2
                 = \int_0^1 dy_1 \cdots dy_4 \; 
                   J_{\rm PS}(\vec{y}) 
                   \left| \mathcal{M} \right|^2
\end{equation}
The different factors are shown in eq.(\ref{eq:phase}).  In the second
step we have rewritten the phase space integral as an integral over
the four--dimensional unit cube, with the appropriate Jacobian.  Like
any integral we can numerically evaluate this phase space integral by
binning the variable we integrate over:
\begin{equation}
\int_0^1 dy \; f(y) 
 \quad \longrightarrow \quad 
 \sum_j (\Delta y)_j f(y_j)
 \sim \Delta y \sum_j f(y_j)
\label{eq:discrete}
\end{equation}
Whenever we talk about numerical integration we can without any loss
of generality assume that the integration boundaries are 0...1. The
integration variable $y$ we can divide into a discrete set of points
$y_j$, for example defined as equi-distant on the $y$ axis or by
choosing some kind of random number $y_j \epsilon [0,1]$. In the
latter case we need to keep track of the bin widths $(\Delta y)_j$. In
a minute, we will discuss how such a random number can be chosen in
more or less smart ways; but before we discuss how to best evaluate
such an integral numerically, let us first illustrate that this
integral is much more useful than just providing the total cross
section. If we are interested in a distribution of an observable, like
for example the distribution of the transverse momentum of a muon
in the Drell--Yan process, we need to compute $d \sigma(p_T)/d
p_T$. This distribution is given by:
\begin{alignat}{5}
\sigma &= \int dy_1 \cdots dy_N \; f(\vec{y}) 
        = \int dy_1 \; \frac{d\sigma}{dy_1} \notag \\
\frac{d\sigma}{dy_1} \Bigg|_{y_1^0} &= \int dy_2 \cdots dy_N \; f(y_1^0)
                             = \int dy_1 \cdots dy_N \; f(\vec{y})  
                                      \; \delta(y_1-y_1^0)
\end{alignat}
We can compute this distribution numerically in two ways. One way would
be to numerically evaluate the $y_2 \cdots y_N$ integrations and just
leave out the $y_1$ integration. The result will be a function of
$y_1$ which we can evaluate at any point $y_1^0$. This method is what
I for example used for Prospino, when I was a graduate student. The
second and much smarter option corresponds to the last term in the
equation above, with the delta distribution defined for discretized
$y_1$. This is not hard to do: first, we define an array the size of
the number of bins in the $y_1$ integration. Then, for each $y_1$
value of the complete $y_1 \cdots y_N$ integration we decide where it
goes in this array and add $f(\vec{y})$ to this array. And finally, we
print $f(y_1)$ to see the distribution. This array is referred to as a
histogram and can be produced for example using the CernLib. This
\underline{histogram approach} does not look like much, but 
imagine you want to compute a distribution $d\sigma/dp_T$, where
$p_T(\vec{y})$ is a complicated function of the integration variables,
so you want to compute:
\begin{equation}
\frac{d\sigma}{dp_T} = \int dy_1 \cdots dy_N \; f(\vec{y})  
                                      \; \delta \left( p_T(\vec{y})-p_T^0 
                                               \right)
\end{equation}
Histograms mean that when we compute the total cross section entirely
numerically we can trivially extract all distributions in the same
process.\bigskip

The procedure outlined above has an interesting interpretation.
Imagine we do the entire phase space integrations numerically. Just
like computing the interesting observables we can compute the momenta
of all external particles. These momenta are not all independent,
because of energy--momentum conservation, but this can be taken care
of. The tool which translates the vector of integration variables
$\vec{y}$ into the external momenta is called a \underline{phase space
generator}.  Because the phase space is not uniquely defined in terms
of the integration variables, the phase space generator also has to
return the Jacobian $J_{\rm PS}$, the phase space weight. If we think
of the integration as an integration over the unit cube, this weight
is combined with the matrix element squared $\left| \mathcal{M}
\right|^2$. Once we compute the unique phase space configuration
$(k_1, k_2, p_1 \cdots p_M)_j$ which corresponds to the vector
$\vec{y}_j$ the combined weight $W = J_{\rm PS} \left| \mathcal{M}
\right|^2$ is simply the probability that this configuration will
appear at the LHC. Which means, we do not only integrate over the
phase space, we really simulate events at the LHC.  The only
complication is that the probability of a certain configuration is not
only given my the frequency with which it appears, but also by the
additional explicit weight. So when we run our numerical integration
through the phase space generator and histogram all the distributions
we are interested in we really generate
\underline{weighted events}. These events, \ie the momenta of all
external particles and the weight $W$, we can for example store in a
big file.\bigskip

This simulation is not quite what experimentalists want --- they want
to represent the probability of a certain configuration appearing only
by its frequency. This means we have to unweight the events and
translate the weight into frequency. To achieve this we normalize all
our event weights to the maximum weight $W_{\rm max}$, \ie compute the
ratio $W_j/W_{\rm max} \epsilon [0,1]$, generate a flatly distributed
random number $r \epsilon [0,1]$, and keep the event if $W_j/W_{\rm
  max} > r$. This guarantees that each event $j$ survives with a
probability $W_j/W_{\rm max}$, which is exactly what we want --- the
higher the weight the more likely the event stays. The challenge in
this translation is only that we will lose events, which means that
our distributions will if anything become more ragged. So if it
weren't for the experimentalists we would never use
\underline{unweighted events}. I should add that experimentalists have
a good reason to want such unweighted events, because they feed best
through their detector simulations. 

The last comment is that if the phase space configuration $(k_1, k_2,
p_1 \cdots p_M)_j$ can be measured, its weight $W_j$ better be
positive. This is not trivial once we go beyond leading order.  There,
we need to add several contributions to produce a physical event, like
for example different $n$--particle final states, and there is no need
for all of them to be positive. All we have to guarantee is that after
adding up all contributions and after integrating over any kind of
unphysical degree of freedom we might have introduced, the probability
of a physics configuration is positive. For example, negative values
for parton densities are not problematic, as long as we always have a
positive hadronic rate $d\sigma_{pp \to X}>0$.\bigskip

The numerical phase space integration for many particles faces two
problems. First, the partonic phase space for $M$ on-shell particles
in the final state has $3(M+2)-3$ dimensions. If we divide each of
these directions in 100 bins, the number of phase space points we need
to evaluate for a $2 \to 4$ process is $100^{15}=10^{30}$, which is
not realistic.

To integrate over a large number of dimensions we use \underline{Monte
Carlo integration}. In this approach we define a distribution $p_Y(y)$
such that for a one-dimensional integral we can replace the binned
discretized integral in eq.(\ref{eq:discrete}) with a discretized
version based on a set of random numbers $Y_j$ over the $y$
integration space
\begin{equation}
\langle g(Y) \rangle
= \int_0^1 dy \; p_Y(y) \; g(y) 
\quad \longrightarrow \quad 
\frac{1}{N} \sum_j g(Y_j)
\end{equation}
All we have to make sure is that the probability of returning $Y_j$ is
given by $p_Y(y)$ for $y < Y_j < y + dy$. This form has the advantage
that we can naively generalize it to any number of $n$ dimensions,
just by organizing the random numbers $Y_j$ in one large vector
instead of an $n$-dimensional array.

Our $n$-dimensional phase space integral listed above we can rewrite
the same way:
\begin{equation}
\int_0^1 d^ny \; f(y) 
  = \int_0^1 d^ny \; \frac{f(y)}{p_Y(y)} \; p_Y(y)
  = \left< \frac{f(Y)}{p_Y(Y)} \right> 
\quad \longrightarrow \quad 
\frac{1}{N} \sum_j \frac{f(Y_j)}{p_Y(Y_j)}
\end{equation}
In other words, we have written the phase space integral in a
discretized way which naively does not involve the number of
dimensions any longer. All we have to do to compute the integral is
average over $N$ phase space values of $f/p_Y$. In the ideal case
where we exactly know the form of the integrand and can map it into
our random numbers, the error of the numerical integration will be
zero. So what we have to find is a way to encode $f(Y_j)$ into
$p_Y(Y_j)$. This task is called \underline{importance sampling} and
you will have to find some documentation for example on Vegas to look
at the details. 

Technically, you will find that Vegas will call the function which
computes the weight $W = J_{\rm PS} |\mathcal{M}|^2$ for a number of
phase space points and average over these points, but including
another weight factor $W_{\rm MC}$ representing the importance
sampling. If you want to extract distributions via histograms you have
to therefore add the total weight $W = W_{\rm MC} J_{\rm PS}
|\mathcal{M}|^2$ to the columns. \bigskip

The second numerical challenge is that the matrix elements for
interesting processes are by no means flat, and we would like to help
our adaptive (importance sampling) Monte Carlo by defining the
integration variables such that the integrand is as flat as possible.
Take for example the integration over the partonic momentum fraction,
where the integrand is usually falling off at least as $1/x$. So we
can substitute
\begin{equation}
\int_\delta dx \; \frac{C}{x} = 
\int_{\log \delta} d \log x \; \left( \frac{d \log x}{dx} \right)^{-1} \; 
                   \frac{C}{x} =
\int_{\log \delta} d \log x \; C 
\end{equation}
and improve our integration significantly. Moving on to a more
relevant example: particularly painful are intermediate particles with
Breit--Wigner propagators squared, which we need to integrate over the
momentum $s = p^2$ flowing through:
\begin{equation}
P(s,m) = \frac{1}{(s-m^2)^2 + m^2 \Gamma^2}
\end{equation}
For example the Standard-Model Higgs boson with a mass of 120~GeV has
a width around $0.005$~GeV, which means that the integration over the
invariant mass of the Higgs decay products $\sqrt{s}$ requires a
relative resolution of $10^{-5}$. Since this is unlikely to be
achievable, what we should really do is find a substitution which
produces the inverse Breit--Wigner as a Jacobian and leads to a flat
integrand --- et voil{\'a}
\begin{alignat}{5}
\int ds \; \frac{C}{(s-m^2)^2 + m^2 \Gamma^2} 
&= \int dz \; \left( \frac{dz}{ds} \right)^{-1} 
              \frac{C}{(s-m^2)^2 + m^2 \Gamma^2} 
\notag \\
&= \int dz \; \frac{(s-m^2)^2 + m^2 \Gamma^2}{m \Gamma} \;
              \frac{C}{(s-m^2)^2 + m^2 \Gamma^2} 
\notag \\
&= \frac{1}{m \Gamma} \int dz \; C 
\qquad \text{with} \qquad \tan z = \frac{s - m^2}{m \Gamma}
\end{alignat}
This is the coolest \underline{phase space mapping} I have seen, and
it is incredibly useful. Of course, an adaptive Monte Carlo will
eventually converge on such an integrand, but a well-chosen set of
integration parameters will speed up our simulations significantly.

\subsection{Helicity amplitudes}

When we compute a transition amplitude, what we usually do is write
down all spinors, polarization vectors, interaction vertices and
propagators and square the amplitude analytically to get
$|\mathcal{M}|^2$. Of course, nobody does gamma--matrix traces by hand
anymore, instead we use powerful tools like Form. But we can do even
better.  As an example, let us consider the simple process $u \bar{u}
\to
\gamma^* \to \mu^+ \mu^-$. The structure of the amplitude in the Dirac
indices involves one vector current on each side $(\bar{u}_f
\gamma_\mu u_f)$. For each $\mu = 0 \cdots 3$ this object gives a
c-number, even though the spinors have four components and each gamma
matrix is a $4 \times 4$ matrix as well. The intermediate photon
propagator has the form $g_{\mu \nu}/s$, which is a simple number as
well and implies a sum over $\mu$ in both of the currents forming the
matrix element.

Instead of squaring this amplitude symbolically we can first compute
it numerically, just inserting the correct numerical values for each
component of each spinor etc, without squaring it. MadGraph is a tool
which automatically produces a Fortran routine which calls the
appropriate functions from the Helas library, to do precisely
that. For our toy process the MadGraph output looks roughly like:
{\scriptsize
\begin{verbatim} 
      REAL*8 FUNCTION UUB_MUPMUM(P,NHEL)
C  
C FUNCTION GENERATED BY MADGRAPH
C RETURNS AMPLITUDE SQUARED SUMMED/AVG OVER COLORS
C FOR PROCESS : u u~ -> mu+ mu-  
C  
      INTEGER    NGRAPHS,    NEIGEN,    NEXTERNAL       
      PARAMETER (NGRAPHS=   1,NEIGEN=  1,NEXTERNAL=4)   
      INTEGER    NWAVEFUNCS     , NCOLOR
      PARAMETER (NWAVEFUNCS=   5, NCOLOR=   1) 

      REAL*8 P(0:3,NEXTERNAL)
      INTEGER NHEL(NEXTERNAL)

      INCLUDE 'coupl.inc'

      DATA Denom(1  )/            1/                                       
      DATA (CF(i,1  ),i=1  ,1  ) /     3/                                  

      CALL IXXXXX(P(0,1   ),ZERO ,NHEL(1   ),+1,W(1,1   ))        
      CALL OXXXXX(P(0,2   ),ZERO ,NHEL(2   ),-1,W(1,2   ))        
      CALL IXXXXX(P(0,3   ),ZERO ,NHEL(3   ),-1,W(1,3   ))        
      CALL OXXXXX(P(0,4   ),ZERO ,NHEL(4   ),+1,W(1,4   ))        
      CALL JIOXXX(W(1,1   ),W(1,2   ),GAU ,ZERO    ,ZERO    ,W(1,5   ))    
      CALL IOVXXX(W(1,3   ),W(1,4   ),W(1,5   ),GAL ,AMP(1   ))            
      JAMP(   1) = +AMP(   1)

      DO I = 1, NCOLOR
          DO J = 1, NCOLOR
              ZTEMP = ZTEMP + CF(J,I)*JAMP(J)
          ENDDO
          UUB_MUPMUM =UUB_MUPMUM+ZTEMP*DCONJG(JAMP(I))/DENOM(I)   
      ENDDO
      END
\end{verbatim}
}
The input to this function are the external momenta and the helicities
of all fermions in the process. Remember that helicity and chirality
are identical only for massless fermions. In general, chirality is
defined as the eigenvalue of the projectors $(\one \pm \gamma_5)/2$,
while helicity is defined as the projection of the spin onto the
momentum direction, or as the left or right handedness. For each point
in phase space and each helicity combination ($\pm 1$ for each
external fermion) MadGraph computes the matrix element using Helas
routines like for example:
\begin{itemize}
\item[$\cdot$] {\tt IXXXXX}($p,m,n_{\rm hel},n_{\rm sf},F$) computes the 
  wave function of a fermion with incoming fermion number, so either
  an incoming fermion or an outgoing anti-fermion. As input it
  requires the 4-momentum, the mass and the helicity of this
  fermion. Moreover, this particle with incoming fermion number can be
  a particle or an anti-particle. This means $n_{\rm fs} = +1$ for the
  incoming $u$ and $n_{\rm sf} = -1$ for the outgoing $\mu^+$, because
  the particles in MadGraph are defined as $u$ and $\mu^-$. The
  fermion wave function output is a complex array $F(1:6)$.

  Its first two entries are the left-chiral part of the fermionic
  spinor, \ie $F(1:2) = (\one -\gamma_5)/2 \; u$ or $F(1:2) = (\one
  -\gamma_5)/2 \; v$ for $n_{\rm sf} = \pm 1$. The entries $F(3:4)$
  are the right-chiral spinor. These four numbers can be computed
  from the 4-momentum, if we know the helicity of the
  particles. Because for massless particles helicity and chirality are
  identical, our massless quarks and leptons will for example have
  only entries $F(1:2)$ for $n_{\rm hel}=-1$ and $F(3:4)$ for $n_{\rm
  hel}=+1$.

  The last two entries contain the 4-momentum in the direction of the
  fermion flow, namely $F(5) = n_{\rm sf} (p(0)+ip(3))$ and $F(6) =
  n_{\rm sf} (p(1)+ip(2))$. The first four entries in this spinor
  correspond to the size of each $\gamma$ matrix, which is usually
  taken into account by computing the trace of the chain of gamma
  matrices.

\item[$\cdot$] {\tt OXXXXX}($p,m,n_{\rm hel},n_{\rm sf},F$) does the 
  same for a fermion with outgoing fermion flow, \ie our incoming
  $\bar{u}$ and our outgoing $\mu^-$. The left-chiral and
  right-chiral components now read $F(1:2) = \bar{u} (\one -
  \gamma_5)/2$ and $F(3:4) = \bar{u} (\one + \gamma_5)/2$, and
  similarly for the spinor $\bar{v}$.  The last two entries are $F(5)
  = n_{\rm sf} (p(0)+ip(3))$ and $F(6) = n_{\rm sf} (p(1)+ip(2))$.

\item[$\cdot$] {\tt JIOXXX}($F_i,F_o,g,m,\Gamma,J_{io}$) computes the 
  (off-shell) current for the vector boson attached to the two
  external fermions $F_i$ and $F_o$. The coupling $g(1:2)$ is a
  complex array with the interaction of the left-chiral and
  right-chiral fermion in the upper and lower index. Obviously, we
  need to know the mass and the width of the intermediate vector
  boson. The output array $J_{io}$ again has six components:
\begin{alignat}{5}
J_{io}(\mu+1)&= - \frac{i}{q^2} \; F_o^T \; \gamma^\mu
                \left(  g(1) \; \frac{\one - \gamma_5}{2} \;
                      + g(2) \; \frac{\one + \gamma_5}{2} \;
                \right) \;
                F_i \notag \\
J_{io}(5) &= -F_i(5) + F_o(5) 
           \sim - p_i(0) + p_o(0) +i \left( -p_i(3) - p_o(3) 
                                     \right) \notag \\
J_{io}(6) &= -F_i(6) + F_o(6)
           \sim - p_i(1) + p_o(1) +i \left( -p_i(2) + p_o(2) 
                                     \right) \notag \\
\Rightarrow \qquad 
q^\mu &= \left( \text{Re} J_{io}(5), \text{Re} J_{io}(6),
                \text{Im} J_{io}(6), \text{Im} J_{io}(5)
        \right) 
\end{alignat}
  The last line illustrates why we need the fifth and sixth arguments
  of $F_{io}$. The first four entries in $J_{io}$ correspond to the
  index $\mu$ in this vector current, while the index $j$ of the
  spinors has been contracted between $F_o^T$ and $F_i$.

\item[$\cdot$] {\tt IOVXXX}($F_i,F_o,J,g,V$) computes the amplitude of a 
  fermion--fermion--vector coupling using the two external fermionic
  spinors $F_i$ and $F_o$ and an incoming vector current $J$. Again,
  the coupling $g(1:2)$ is a complex array, so we numerically compute 
\begin{alignat}{5}
   F_o^T \; \sla{J} \left( g(1) \; \frac{\one - \gamma_5}{2} \;
                         + g(2) \; \frac{\one + \gamma_5}{2} \;
                    \right) \; F_i
\end{alignat}
  We see that all indices $j$ and $\mu$ of the three input arguments
  are contracted in the final result. Momentum conservation is not
  explicitly enforced by {\tt IOVXXX}, so we have to take care of it
  beforehand.
\end{itemize}

Given the list above it is easy to see how MadGraph computes the
amplitude for $u \bar{u} \to \gamma^* \to \mu^+ \mu^-$. First, it
always calls the wave functions for all external particles and puts
them into the array $W(1:6,1:4)$. The vectors $W(*,1)$ and $W(*,3)$
correspond to $F_i(u)$ and $F_i(\mu^+)$, while $W(*,2)$ and $W(*,4)$
mean $F_o(\bar{u})$ and $F_o(\mu^-)$. The first vertex we evaluate is
the $\bar{u} \gamma u$ vertex, which given $F_i = W(*,1)$ and $F_o =
W(*,2)$ uses {\tt JIOXXX} to compute the vector current for the
massless photon in the $s$ channel. Not much would change if we
instead chose a massive $Z$ boson, except for the arguments $m$ and
$\Gamma$ in the {\tt JIOXXX} call. The {\tt JIOXXX} output is the
photon current $J_{io} \equiv W(*,5)$. The second step combines this
current with the two outgoing muons in the $\mu^+ \gamma \mu^-$
vertex. Since this number gives the final amplitude, it should return
a c-number, no array. MadGraph calls {\tt IOVXXX} with $F_i = W(*,3)$
and $F_o = W(*,4)$, combined with the photon current $J = W(*,5)$. The
result {\tt AMP} is copied into {\tt JAMP} without an additional sign
which could have come from the ordering of external fermions. The only
remaining sum left to compute before we square {\tt JAMP} is the color
structure, which in our simple case means one color structure with a
color factor $N_c=3$.\bigskip

Of course, to calculate the transition amplitude MadGraph requires all
masses and couplings. They are transferred through common blocks in
the file coupl.inc and computed elsewhere.  In general, MadGraph uses
unitary gauge for massive vector bosons, because in the
helicity amplitude approach it is easy to accommodate complicated
tensors, in exchange for a large number of Feynman diagrams.\bigskip

The function {\tt UUB\_MUPMUM} described above is not yet the full
story. Remember that when we square $\mathcal{M}$ symbolically we need
to sum over the spins of the outgoing states to transform a spinor
product of the kind $u \bar{u}$ into the residue or numerator of a
fermion propagator. To obtain the final result numerically we also
need to sum over all possible helicity combinations of the external
fermions, in our case $2^4 = 16$ combinations.
{\scriptsize
\begin{verbatim} 

      SUBROUTINE SUUB_MUPMUM(P1,ANS)
C  
C FUNCTION GENERATED BY MADGRAPH
C RETURNS AMPLITUDE SQUARED SUMMED/AVG OVER COLORS
C AND HELICITIES FOR THE POINT IN PHASE SPACE P(0:3,NEXTERNAL)
C  
C FOR PROCESS : u u~ -> mu+ mu-  
C  
      INTEGER    NEXTERNAL,   NCOMB,       
      PARAMETER (NEXTERNAL=4, NCOMB= 16)
      INTEGER    THEL
      PARAMETER (THEL=NCOMB*1)

      REAL*8 P1(0:3,NEXTERNAL),ANS

      INTEGER NHEL(NEXTERNAL,NCOMB),NTRY
      REAL*8 T, UUB_MUPMUM
      INTEGER IHEL,IDEN,IC(NEXTERNAL)
      INTEGER IPROC,JC(NEXTERNAL)
      LOGICAL GOODHEL(NCOMB)

      DATA GOODHEL/THEL*.FALSE./
      DATA NTRY/0/

      DATA (NHEL(IHEL,  1),IHEL=1,4) / -1, -1, -1, -1/
      DATA (NHEL(IHEL,  2),IHEL=1,4) / -1, -1, -1,  1/
      DATA (NHEL(IHEL,  3),IHEL=1,4) / -1, -1,  1, -1/
      DATA (NHEL(IHEL,  4),IHEL=1,4) / -1, -1,  1,  1/
      DATA (NHEL(IHEL,  5),IHEL=1,4) / -1,  1, -1, -1/
      DATA (NHEL(IHEL,  6),IHEL=1,4) / -1,  1, -1,  1/
      DATA (NHEL(IHEL,  7),IHEL=1,4) / -1,  1,  1, -1/
      DATA (NHEL(IHEL,  8),IHEL=1,4) / -1,  1,  1,  1/
      DATA (NHEL(IHEL,  9),IHEL=1,4) /  1, -1, -1, -1/
      DATA (NHEL(IHEL, 10),IHEL=1,4) /  1, -1, -1,  1/
      DATA (NHEL(IHEL, 11),IHEL=1,4) /  1, -1,  1, -1/
      DATA (NHEL(IHEL, 12),IHEL=1,4) /  1, -1,  1,  1/
      DATA (NHEL(IHEL, 13),IHEL=1,4) /  1,  1, -1, -1/
      DATA (NHEL(IHEL, 14),IHEL=1,4) /  1,  1, -1,  1/
      DATA (NHEL(IHEL, 15),IHEL=1,4) /  1,  1,  1, -1/
      DATA (NHEL(IHEL, 16),IHEL=1,4) /  1,  1,  1,  1/
      DATA (  IC(IHEL,  1),IHEL=1,4) /  1,  2,  3,  4/
      DATA (IDEN(IHEL),IHEL=  1,  1) /  36/

      NTRY=NTRY+1

      DO IHEL=1,NEXTERNAL
         JC(IHEL) = +1
      ENDDO

      DO IHEL=1,NCOMB
         IF (GOODHEL(IHEL,IPROC) .OR. NTRY .LT. 2) THEN
            T = UUB_MUPMUM(P1,NHEL(1,IHEL),JC(1))            
            ANS = ANS + T
            IF (T .GT. 0D0 .AND. .NOT. GOODHEL(IHEL,IPROC)) THEN
               GOODHEL(IHEL,IPROC)=.TRUE.
            ENDIF
         ENDIF
      ENDDO
      ANS = ANS/DBLE(IDEN)

      END
\end{verbatim}
}
The important part of this subroutine is the list of possible helicity
combinations stored in the array {\tt NHEL}$(1:4,1:16)$. Adding all
different helicity combinations (of which some might well be zero)
means a loop over the second argument and a call of {\tt UUB\_MUPMUM}
with the respective helicity combination. The complete spin--color
averaging factor is included as {\tt IDEN} and given by $2 \times 2
\times N_c^2 = 36$. So MadGraph indeed provides us with a subroutine
{\tt SUUB\_MUPMUM} which numerically computes
$\overline{|\mathcal{M}|^2}$ for each phase space point, \ie external
momentum configuration. MadGraph also produces a file with all Feynman
diagrams contributing to the given subprocess, in which the numbering
of the external particles corresponds to the second argument of $W$
and the argument of {\tt AMP} is the numbering of the Feynman
diagrams. After looking into the code very briefly we can also easily
identify different intermediate results $W$ which will only be
computed once, even if they appear several times in the different
Feynman diagrams.

The helicity method might not seem particularly appealing for a simple
$2 \to 2$ process, but it makes it easily possible to compute
processes with four and more particles in the final state and up to
10000 Feynman diagrams which we could never square symbolically, no
matter how many graduate students' lives we turn into hell.

\subsection{Errors}

As argued in the very beginning of the lecture, LHC physics always
means extracting signals from often large backgrounds. This means, a
correct error estimate is crucial. For LHC calculations we are usually
confronted with three types of errors.\bigskip

The first and easiest one are the \underline{statistical errors}. For
small numbers of events these experimental errors are described by
Poisson statistics, and for large numbers they converge to the
Gaussian limit. And that is about the only complication we encounter
for them.\bigskip

The second set of errors are \underline{systematic errors}, like for
example the calibration of the jet and lepton energy scales, the
measurements of the luminosity, or the efficiencies to identify a muon
as a muon. Some of you might remember what happened last, when a bunch
of theorists mistook a forward pion for an electron --- that happened
right around my TASI, and people had not only discovered
supersymmetry, but also identified its breaking mechanism. Of course,
our experimentalist CDF lecturer told us immediately that the whole
thing was a joke. Naively, we would not assume that systematic are
Gaussian, but remember that we determine these numbers largely from
well-understood background processes. Such counting experiments in
background channels like $Z
\to$~leptons, however, do behave Gaussian. The only caveat is the
shape of far-away tails, which can turn out to be bigger than the
exponentially suppressed Gaussian shape. \bigskip

The last source of errors are \underline{theory errors}, and they are
hardest to model, because they are dominated by higher-order QCD
effects, fixed order or enhanced by large logarithms. If we could
compute all remaining higher-order terms, we would do so, which means
everything else is a wild guess. Moreover, higher-order effects are
not any more likely to give a relative $K$ factor of 1.0 than 0.9 or
1.1. In other words, theory errors cannot have a peak and they are
definitely not Gaussian. There is a good reason to choose the Gaussian
short cut, because we know that folding three Gaussian errors gives us
another Gaussian error, which makes things so much easier. But this
lazy approach assumes the we know much more about QCD than we actually
do, so please stop lying. On the other hand, we also know that theory
errors cannot be arbitrarily large. Unless there is a very good
reason, a $K$ factor for a total LHC cross section should not be
larger than something like 3. If that were the case, we would conclude
that perturbative QCD breaks down, and the proper description of error
bars would be our smallest problem. In other words, the centrally flat
theory probability distribution for an LHC observable has to go to
zero for very large deviations from the currently best value.

A good solution to this problem is the so-called \underline{Rfit
scheme}, used for example by the CKMfitter or the SFitter
collaborations. It starts from the assumption that for very large
deviations there will always be tails from the experimental errors, so
we can neglect the impact of the theory errors on this range. In the
center of the distribution we simply cut open the experimental
Gaussian-type distribution and insert a flat theory piece. We could
also modify the transition region by changing for example the width of
the experimental Gaussian error as an effect of a falling-off theory
error, but in the simplest model we just use a log-likelihood $\chi^2
= -2 \log
\mathcal{L}$ given a set of measurements $\vec d$ and in 
the presence of a general correlation matrix $C$
\begin{alignat}{7}
\chi^2     &= {\vec{\chi}_d}^T \; C^{-1} \; \vec{\chi}_d  \notag \\ 
\chi_{d,i} &=
  \begin{cases}
  0  
          &|d_i-\bar{d}_i | <   \sigma^{\text{(theo)}}_i \\
  \frac{\D d_i-\bar{d}_i+ \sigma^{\text{(theo)}}_i}{\D \sigma^{\text{(exp)}}_i}
          & \; d_i-\bar{d}_i\; < - \sigma^{\text{(theo)}}_i \\
  \frac{\D d_i-\bar{d}_i- \sigma^{\text{(theo)}}_i}{\D \sigma^{\text{(exp)}}_i}
  \qquad  & \; d_i-\bar{d}_i\; >   \sigma^{\text{(theo)}}_i \; .
  \end{cases}
\label{eq:flat_errors}
\end{alignat}
And that is it, all three sources of LHC errors can be described
correctly, and nothing stops us from computing likelihood maps to
measure the top mass or identify new physics or just have some fun in
life at the expense of the Grid.

\section*{Further reading, acknowledgments, etc.}

This is the point where the week in beautiful Boulder is over and I
should thank K.T. and his Boulder team as well as our two organizers
for their kind invitation. I typed most of these notes in Boulder's
many nice cafes and 11 years after I went here as a student TASI and
Boulder still make the most enjoyable and most productive school in
our field. Whoever might ever think about moving it away from Boulder
cannot possibly have the success of the school in mind.

It has been great fun, even though QCD has a reputation of being a dry
topic. I hope you enjoyed learning it as much as I enjoyed learning it
while teaching it. Just like most of you I am really only a QCD user,
but for an LHC phenomenologists there is no excuse for not knowing the
relevant aspects of QCD. Have fun in the remaining lectures, write
some nice theses, and I hope I will see as many of you as possible
over the coming 20 years.
%
%
%
LHC physics need all the help we can
get, and it is great fun, so please come and join us!\bigskip \bigskip

Of course there are many people I need to thank for helping me write
these notes: Fabio Maltoni, Johan Alwall and Steffen Schumann for
having endured a great number of critical questions and for convincing
me that jet merging is the future; Steffen Schumann, Ben Allanach and
Tom DeGrand for their comments on this draft; Beate Heinemann for
providing me with one of the most interesting plots from the Tevatron
and for answering many stupid questions over the years --- as did Dirk
Zerwas and Kyle Cranmer.\bigskip \bigskip

You note that this writeup, just like the lectures, is more of an
informal chat about LHC physics than a proper review paper. But if I
had not cut as many corners we would never have made it to the fun
topics.  In the same spirit, there is no point in giving you a list of
proper original references, so I would rather list a few books and
review articles which might come in handy if you would like to know
more:

\begin{itemize}
\item[--] I started learning high--energy theory including QCD from
  Otto Nachmann's book. I still use his appendices to look up Feynman
  rules, because I have rarely seen another book with as few (if not
  zero) typos~\cite{Nachtmann:1990ta}. Similar, but maybe a little
  more modern is the primer by Cliff Burgess and Guy
  Moore~\cite{Burgess:2007zi}. At the end of it you will find more
  literature tips.
  
\item[--] For a more specialized book on QCD have a look at the pink
  book by Ellis, Stirling, Webber. It includes everything you ever
  wanted to know about QCD~\cite{Ellis:1991qj}. Maybe a little more
  phenomenology you can find in G\"unther Dissertori, Ian Knowles and
  Michael Schmelling's book on QCD and
  phenomenology~\cite{Dissertori:2003pj}.
  
\item[--] If you would like to learn how to for example compute
  higher-order cross sections to Drell--Yan production, Rick Field
  works it all out in his book~\cite{Field:1989uq}.

\item[--] Unfortunately, there is comparably little literature on jet
  merging yet. The only review I know is by Michelangelo Mangano and
  Tim Stelzer~\cite{Mangano:2005dj}. There is a very concise
  discussion included with the comparison of the different
  models~\cite{Alwall:2007fs}. If you want to know more, you will have
  to consider the original literature or wait for the review article
  which Frank Krauss and Peter Richardson promised to write for
  Journal of Physics G.
  
\item[--] Recently, I ran across George Sterman's TASI lectures. They
  are comparably formal, but they are a great read if you know
  something about QCD already~\cite{Sterman:2004pd}.
  
\item[--] For MC\@@NLO there is nothing like the original papers. Have
  a look at Bryan Webber's and Stefano Frixione's work and you cannot
  but understand what it is about~\cite{Frixione:2002ik}!

\item[--] For CKKW, look at the original paper.  It beautifully
  explains the general idea on a few pages, at least for final state
  radiation~\cite{Catani:2001cc}.

\item[--] If you are using Madgraph to compute helicity
  amplitudes there is the original bright green documentation which
  describes every routine in detail. You might want to check the
  format of the arrays, if you use for example the updated version
  inside MadEvent~\cite{Murayama:1992gi}.

\end{itemize}

\end{document}